\documentclass[10pt,twocolumn,showpacs,preprintnumbers,amsmath,amssymb,aps,prl,superscriptaddress,longbibliography]{revtex4-2}
\usepackage{appendix}
\usepackage{bbm}
\usepackage{mathrsfs}
\usepackage{graphicx}
\usepackage{dcolumn}
\usepackage{bm}
\usepackage{braket}
\usepackage{amsmath}
\usepackage{amsfonts}
\usepackage[dvipsnames]{xcolor}
\usepackage[colorlinks=true,linkcolor=Blue,urlcolor=BlueViolet,citecolor=BlueViolet]{hyperref}
\usepackage{natbib}
\usepackage{multirow}

\begin{document}
	
\title{Layer-dependent quantum anomalous Hall effect in rhombohedral graphene}
\author{Zhaochen Liu}
\affiliation{State Key Laboratory of Surface Physics and Department of Physics, Fudan University, Shanghai 200433, China}
\affiliation{Shanghai Research Center for Quantum Sciences, Shanghai 201315, China}
\author{Jing Wang}
\thanks{Contact author: wjingphys@fudan.edu.cn}
\affiliation{State Key Laboratory of Surface Physics and Department of Physics, Fudan University, Shanghai 200433, China}
\affiliation{Shanghai Research Center for Quantum Sciences, Shanghai 201315, China}
\affiliation{Institute for Nanoelectronic Devices and Quantum Computing, Fudan University, Shanghai 200433, China}
\affiliation{Hefei National Laboratory, Hefei 230088, China}
	
\begin{abstract}
The quantum anomalous Hall (QAH) effect, first proposed in the Haldane model, is a paradigmatic example of the application of band topology in condensed matter physics. The recent experimental discoveries of high Chern number QAH effect in pentalayer and tetralayer rhombohedral graphene highlight the intriguing interplay between strong interactions and spin-orbit coupling (SOC). Here we propose a minimal interacting model for spin-orbit-coupled rhombohedral graphene and use the Hartree-Fock analysis to explore the phase diagram at charge neutrality. We find that with Ising SOC on one outmost graphene layer, the in-plane layer-antiferromagnetic order is the insulating ground state without displacement field. Upon increasing the gate displacement field, we find that the QAH state with Chern number being equal to the layer number emerges between layer-antiferromagnetic state and layer-polarized state, which is consistent with experimental observations. Remarkably, we study the phase diagram for different thicknesses and 
find pentalayer is optimal for the QAH effect. Finally, we propose that the QAH state is enlarged by engineering opposite Ising SOC on the opposite outmost layers of rhombohedral graphene. These results will facilitate the realization of QAH states in rhombohedral graphene with different thicknesses. Our work serves as a foundation for further exploration of correlated physics of insulating state in rhombohedral graphene.
\end{abstract}
	
	
\maketitle
	
The quantum anomalous Hall (QAH) effect, a prominent outcome of band topology, has attracted significant attention in condensed 
matter physics~\cite{chang2023,tokura2019,wang2017c}. The manifestation of the QAH effect necessitates the breaking of time-reversal symmetry. 
For the QAH effect discovered in magnetic topological insulators, such symmetry breaking originates from the long-range magnetic ordering 
of local moments~\cite{chang2013,checkelsky2014,kou2014,bestwick2015,mogi2015,deng2020}. In the renowned Haldane model, 
the time-reversal breaking arises from a complex hopping term induced by the staggered magnetic flux~\cite{haldane1988}. 
In practical scenarios, such a complex hopping can be generated by electron interaction or spin-orbit coupling (SOC)~\cite{kane2005,raghu2008}. 
However, for monolayer graphene, both the interaction and SOC are too weak to actualize the QAH effect. 
The moir\'e system could overcome this limitation, for electron interactions dominate over the quenched kinetic energy, 
exemplified by the discovery of integer and fractional Chern insulators~
\cite{spanton2018,serlin2020,chen2020,xie2021,li2021,zhao2024,Cai2023,Zeng2023,Park2023,xu2023,lu2024,regnault2011}.
Moreover, the high Chern number QAH states are exceptionally rare in nature. Their realization could pave the way for the
development of next-generation, low-dissipation electronic devices~\cite{Zhao2020,Wang2013b}.

Meanwhile, correlated insulators and the QAH effect have been observed in ABC-stacked rhombohedral multilayer graphene (RMG) without 
any moir\'e patten~\cite{myhro2018,shi2020,kerelsky2021,han2024a,liu2024,zhou2023,han2023b,sha2023,zhou2024}. Near charge neutrality, 
the low-energy electrons of RMG are characterized by an isospin index that includes valley and spin~\cite{grueneis2008,koshino2009a,zhang2010a,jung2013a}. 
The bulk orbitals are dimerized, akin to the Su-Schrieffer-Heeger chain. The wavefunctions of the two bands closest to the Fermi level 
reside mostly on the non-dimerized sites in the outmost layers, with a $k^N$ energy dispersion for $N$ layers at each valley. 
These flat bands have large density of states and carry a large valley-dependent Berry phase, which results in correlated and topological states. 
For instance, superconductivity is observed in doped RMG~\cite{zhou2021,zhou2021a,zhou2022,pantaleon2023,zhang2023}. 
At charge neutrality, an energy gap is spontaneously generated by Coulomb interaction
~\cite{min2008,vafek2010,lemonik2010,nandkishore2010,zhang2011,jung2011,scherer2012,pamuk2017}. Theoretically, these competing symmetry-breaking gapped 
states exhibit quantized anomalous, spin or valley Hall effects, which can be classified according to their Hall conductivity
~\cite{nandkishore2010,zhang2011,jung2011}. Experimentally, an insulating ground state without any flavor Hall effect was found 
at charge neutrality in rhombohedral tetralayer and pentalyer graphene~\cite{kerelsky2021,han2024a,liu2024,zhou2023}. 
This correlated state is expected to have layer antiferromagnetic (LAF) order, where its band gap continuously decreases 
and eventually transitions to the layer-polarized (LP) state as the gate displacement field increases. Intriguingly, 
the QAH state emerges between the LAF and LP state in the presence of Ising SOC from nearby transition metal dichalcogenide 
(TMD)~\cite{gmitra2015,gmitra2017,li2019,naimer2021,zollner2023}, where Chern number is $\mathcal{C}=4$ and $5$ for 
tetralayer and pentalyer~\cite{han2023b,sha2023}, respectively. These remarkable observations naturally lead to important questions. 
What is the \emph{minimal} microscopic interacting model for describing the phase diagram? How is the layer dependence of the QAH effect in RMG? 
Why zero-field QAH effect was observed only at negative displacement field in pentalayer~\cite{han2023b}? 
How to enlarge the QAH state in the phase space?

In this Letter, inspired by these experimental observations~\cite{han2023b,sha2023}, we answer these questions by proposing a minimal interacting model for proximitized RMG, and investigating the phase diagram at charge neutrality using self-consistent Hartree-Fock calculations. The interactions include the short-range repulsive density-density interaction and ferromagnetic inter-valley Hund's 
coupling~\cite{you2022,lu2022,chatterjee2022,zhumagulov2023,zhumagulov2023a,koh2024,supple}. The interaction strength was determined to be consistent with the experimental results. We study the phase diagram for different thicknesses and find the pentalayer is optimal for the QAH effect, which explains well the experiments. The good agreement between our calculation and experimental results strongly validates our model. We further propose that the QAH state is enlarged by engineering opposite Ising SOC on the top and bottom layers of RMG.

\begin{figure}[t]  
\begin{center}
\includegraphics[width=3.4in, clip=true]{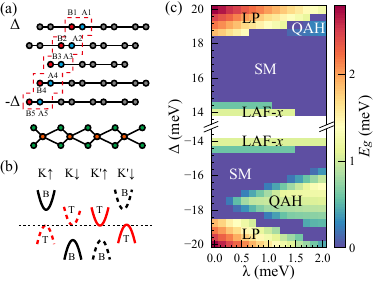}
\end{center}
\caption{(a) The sideview of lattice structure of rhombohedral pentalayer graphene neighboring the top of monolayer TMD. The red dashed box shows the unit-cell, with ($B_i, A_{i+1}$) being strongly hybridized ($i=1,2,3,4$) and non-dimerized sublattices $A_1/B_5$. (b) Schematic of non-interacting band structure of four flavors
  with finite Ising SOC and vanishing inter-layer potential $\Delta=0$. Red and black colors of each isopsin band structure correspond to the top and bottom graphene layers, respectively. Solid and dashed curves correspond to $-5/2$ and $+5/2$ Chern number, respectively. (c) Hartree-Fock phase diagram of pentalayer graphene at charge neutrality, showing the band gap $E_g$ as a function of Ising SOC $\lambda$ and surface potential. All other parameters $U=40$~\text{eV}$\cdot\mathcal{A}_u$, $V=-U/5$ and $J_{\text{H}}=10$~\text{eV}$\cdot\mathcal{A}_u$.
  LAF-$x$ refers to the in-plane layer antiferromagnetic state, SM refers to the semi-metal state, LP refers to the layer-polarized state, and QAH refers to the quantum anomalous Hall state.}
\label{fig1}
\end{figure} 

\emph{Model.} The system we consider is shown in Fig.~\ref{fig1}(a), where a gate-tunable RMG is in proximity with bottom TMD. The Hamiltonian of the system is
\begin{equation}
\mathcal{H} = \mathcal{H}_0+\mathcal{H}_{\text{soc}}+\mathcal{H}_\text{int}.
\end{equation}
$\mathcal{H}_0$ is the single particle tight-binding Hamiltonian~\cite{grueneis2008,koshino2009a,zhang2010a,jung2013a} of the $N$-layer rhombohedral graphene
\begin{equation}
\mathcal{H}_0=\sum_{\mathbf{k},\tau,s}c^\dagger_{\mathbf{k},\tau,s}h_{\tau}(\mathbf{k})c_{\mathbf{k},\tau,s},
\end{equation}
where $\tau=\pm 1$ represents $K$ and $K'$ valley respectively, $s$ is spin index, and $c_{\mathbf{k},\tau,s}=(c_{A_1,\mathbf{k},\tau,s},c_{B_1,\mathbf{k},\tau,s},\cdots,c_{B_N,\mathbf{k},\tau,s})^T$ is the $2N$-component spinor including all the sublattices, with $A_j$ and $B_j$ denoting the sublattices in $j$-th layer. Taking the tetralayer system as an example, the Hamiltonian is
\begin{equation}
  h_{\tau,\text{4L}}=\begin{pmatrix}
      h_0(1) & h_1 & h_2 & 0\\ 
      h^\dagger_1 & h_0(2) &h_1 & h_2 \\
      h^\dagger_2 & h^\dagger_1 & h_0(3) & h_1 \\
      0 & h^\dagger_2 & h^\dagger_1 & h_0(4) \\
     \end{pmatrix},
     \label{eq:ham_ABC}
\end{equation}
where
\begin{eqnarray}
  h_{\tau,0}(i) &=& \begin{pmatrix}
      \delta(2i-1) & v_0\pi^\dagger \\ 
      v_0 \pi  &\delta(2i)\\ 
\end{pmatrix}, \quad
  h_{\tau,1}=\begin{pmatrix}
      v_4\pi^\dagger & v_3\pi \\ 
      \gamma_1  &v_4\pi^\dagger\\ 
  \end{pmatrix},
  \nonumber\\
  h_{\tau,2}&=&\begin{pmatrix}
      0 & \gamma_2/2\\ 
      0 & 0\\ 
  \end{pmatrix}.
  \label{eq:ham_ABC_block}
\end{eqnarray}
and $\pi=\tau k_x+ik_y$, $\tau=\pm 1$ for $K$ and $K'$ valley. The $\delta(i)=\delta$ for $i=1$ or $2N$ and equals to zero for others. And $v_i=\sqrt{3}a\gamma_i/2$, with $a=0.246$~nm as the lattice constant of graphene. The parameters are chosen from Refs.~\cite{zhou2021,zibrov2018,ghazaryan2023} with $(\gamma_0,\gamma_1,\gamma_2,\gamma_3,\gamma_4,\delta)=(3.1,0.38,-0.015,-0.29,-0.141,0.0105)~\text{eV}$. Explicitly the low energy effective $\mathbf{k}\cdot\mathbf{p}$ model of 
$\mathcal{H}_0$ is $\mathcal{H}_N(\mathbf{k})\propto k^N\left[\cos(N\varphi_\mathbf{k})\varrho_x+\sin(N\varphi_\mathbf{k})\varrho_y\right]$, 
where $\cos\varphi_\mathbf{k}=\tau k_x/k$ and $\sin\varphi_\mathbf{k}=k_y/k$, and $\boldsymbol{\varrho}$ 
are Pauli matrices acting on layer pseudospin degree of freedom ($A_1/B_N$). The gate displacement field $\mathcal{D}$ would introduce a interlayer potential between top ($\Delta$) and bottom ($-\Delta$) layers, which would generate a finite gap, as the low energy conduction and valance bands resides mostly in the outmost layers. $\Delta\equiv e\mathcal{D}d/(2\epsilon_r)$, where $e$ is the electric charge, $d$ is the thickness of RMG, and $\epsilon_r\approx 4.5$ is the dielectric constant of hBN substrate~\cite{koh2024}.

The nearby TMD breaks inversion symmetry and introduces Ising SOC in the bottom graphene layer through proximity effect, which breaks the spin $SU(2)$ symmetry to $U(1)$. The previous \emph{ab initio} calculations show that only the neighboring graphene layer could feel the proximitized Ising SOC, thus 
\begin{equation}
\mathcal{H}_{\text{soc}}=\lambda\sum_{\mathbf{k},\tau,s=\pm 1} \tau sc^\dagger_{\eta_N,\mathbf{k},\tau,s}c_{\eta_N,\mathbf{k},\tau,s},
\end{equation}
where $c_{\eta_N,\mathbf{k},\tau,s}$ is electron operator on the $N$-th layer with $\eta=A/B$ for sublattices. In terms of the low energy effective model, the SOC term can be written as $\lambda\tau_zs_z (\varrho_0-\varrho_z)/2$. The Rashba SOC couples two sublattices on the same layer and is negligible for the low energy physics, since a $N$-step process requires electrons hopping between low-energy sites in $A_1$ and $B_N$ via high-energy states.

The interacting Hamiltonian is modeled by the short range repulsive interactions~\cite{you2022,lu2022,chatterjee2022,zhumagulov2023,zhumagulov2023a}, 
\begin{equation}\label{H_int}
\mathcal{H}_{\text{int}}=\int d^2 x\sum_{i}(\frac{U}{2}n_i^2+Vn_{i,K}n_{i,K'}-J_{\text{H}}\mathbf{S}_{i,K}\cdot \mathbf{S}_{i,K'})
\end{equation}
where the summation runs over all sublattices $i$, $n_i\equiv\sum_{\tau,s}n_{i,\tau,s}$, and $n_{i,\tau,s}$ is the electron density operator for $i$-th sublattice, $s$ spin and $\tau$ valley. The local spin density operator is $\mathbf{S}_{i,\tau}=\sum_{s,s'}c^\dagger_{i,\tau,s} \mathbf{s} c_{i,\tau,s'}$. The first term is the spin-valley $SU(4)$ invariant repulsive interaction, and $V$ represents the difference between the intra-valley and inter-valley repulsions which breaks $SU(4)$ symmetry into $SO(4)\simeq SU(2)^{K}_s\times SU(2)^{K'}_s$, and the last term is the Hund's coupling which further reduce the symmetry into total spin $SU(2)_s$ symmetry. We restrict our model to local onsite terms due to large separation between the low-energy sites $A_1$ and $B_N$. As a result, it is justified to exclude the interlayer and inter-sublattice interactions. The Hamiltonian was solved by the self-consistent Hartree-Fock methods. For numerical calculations, we choose $U=40~\text{eV}\cdot\mathcal{A}_u$, $V=-U/5$ and the momentum cut-off $\Lambda=0.16/a_0$, with $a_0$ the lattice constant of graphene and $\mathcal{A}_u$ as area of the unit cell. More numerical  results for other parameters are listed in Supplementary Material~\cite{supple}.

\begin{figure*}[t]
\begin{center}
\includegraphics[width=6.9in, clip=true]{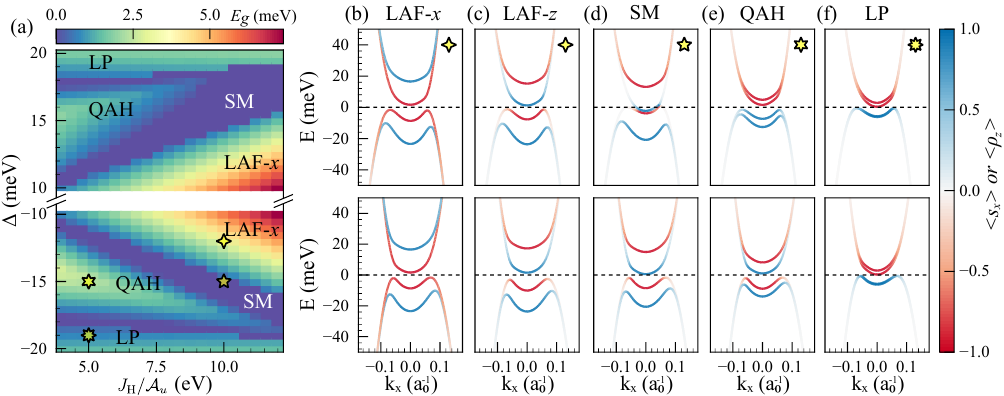}
\end{center}
\caption{(a) Hartree-Fock phase diagram of pentalayer graphene, showing band gap $E_g$ vs $\Delta$ and $J_{\text{H}}$, here $\lambda=1$~meV. (b)-(f) The top and bottom panels show the Hartee-Fock band structure of various phases in (a) around $K$ and $K'$ valley, respectively. (b) \& (c) 1D cut of the Hartree-Fock band structure for LAF-$x$ and LAF-$z$ state at four-pointed star in (a) ($J_{\text{H}}=10~\text{eV}\cdot\mathcal{A}_u$, $\Delta=-12$~meV), respectively. (d) Band structure of SM at pentagram ($J_{\text{H}}=10~\text{eV}\cdot\mathcal{A}_u$, $\Delta=-15$~meV). (e) Band structure of QAH state at hexagram ($J_{\text{H}}=5~\text{eV}\cdot\mathcal{A}_u$, $\Delta=-15$~meV). (d) Band structure of LP insulating state at octagram ($J_{\text{H}}=5~\text{eV}\cdot\mathcal{A}_u$, $\Delta=-19$~meV). The color in (b) represents the spin polarization along $x$-direction $\langle s_x\rangle$. In (c)-(f), the color denotes the layer polarization $\langle\varrho_z\rangle$.}
\label{fig2}
\end{figure*}

\emph{Analysis.} We start by analyzing all possible broken symmetry states before presenting the numerical results. By focusing on the low-energy physics, these broken symmetry states are characterized by the mass term (i.e., the potential difference between $A_1$ and $B_N$) chosen separately for ($K\uparrow$), ($K\downarrow$), ($K'\uparrow$), and ($K'\downarrow$) flavors~\cite{nandkishore2010,zhang2011,jung2011}. Without breaking spin $SU(2)_s$ and valley $U(1)_v$ symmetries, there are $16$ distinct states depending on the sign of the mass term, $m\varrho_z$. We categorize these 16 states into three distinct groups according to their net layer polarization, which is proportional to the sum over the spin valley of the sign of $m$. The first group exhibits full layer polarization (LP), characterized by the four flavors has the same sign of mass term. The second group has partial layer polarization (QAH with $\mathcal{C}=N$) with the sign of one flavor is opposite that of the others. And the third group has zero polarization and is likely to be the ground state in the absence of a gate displacement field, which includes the QAH state ($\mathcal{C}=2N$) with mass term $\tau_z\varrho_z$, the LAF state with mass term $s_z\varrho_z$ and the quantum spin Hall (QSH) state with mass $\tau_zs_z\varrho_z$.

For the interacting model in Eq.~(\ref{H_int}), when $V=0$ and $J_{\text{H}}=0$, all the three states without layer polarization are degenerate. While the presence of Ising SOC, which contains the QSH mass, may trigger the imbalance into the QSH ground state with helical edge states. However, experimentally, only the trivial correlated insulating state was found with and without SOC when the external potential $\Delta$ equals zero~\cite{sha2023,han2024a,han2023b,liu2024}, which indicates an LAF ground state and a finite ferromagnetic Hund's coupling ($J_\text{H}>0$). With Ising SOC, two specific LAF states are possible, namely, the out-of-plane ordered LAF-$z$ and in-plane ordered LAF-$x$. As we noticed that the Ising SOC term, denoted as $\lambda\tau_z s_z(\varrho_0-\varrho_z)/2$, not only introduces a gap opening mass term, but also includes a flavor dependent chemical potential. Therefore, as illustrated in Fig.~\ref{fig1}(b), a non-trivial band edge alignment was realized. For example, the valence band edge of $(K,\uparrow)$ valley is aligned with the 
conduction band edge of $(K,\downarrow)$ and $(K',\uparrow)$ valleys, where both of them are polarized to the top layer. Then two states are possible ground state, one is planar LAF-$x$ with mass $\tau_zs_{x/y}\varrho_z$, and the other is inter-valley coherence state with mass $\tau_{x/y}s_{z}\varrho_z$. Numerically, we found that LAF-$x$ is favorable.

\emph{Phase diagram.} We now explore the phase diagram by tuning $\lambda$, $\Delta$, and $J_{\text{H}}$. Experimentally, the proximate Ising SOC strength, on the order of $1$~meV, can be varied by changing the TMD materials and also by adjusting the twisting angle between graphene and TMD\cite{li2019,naimer2021,zollner2023}. Fig.~\ref{fig1}(c) is the Hartree-Fock phase diagram 
of pentalayer graphene, showing the band gap $E_g$ with respect to $\lambda$ and $\Delta$, for $J_{\text{H}}=10~\text{eV}\cdot\mathcal{A}_u$. For $\Delta$ smaller than the critical value of the LAF state to semi-metal 
transition (about $15$~meV), the system exhibits LAF-$x$ order. As $\Delta$ increases, a weak first-order phase transition from the LAF-$x$ state to the LAF-$z$ state occurs where the band gap changes about $0.5$~meV~\cite{supple}. The phase transition happens when LAF-$z$ is semi-metallic (SM), while LAF-$x$ still maintains a tiny gap. The SM state is a compensated semi-metal with electron and hole pockets. Meanwhile, the gap of LAF-$x$ state is independent of the SOC strength, which is in agreement with the above band structure analysis. The critical potential $\Delta$ for the LAF to SM transition, corresponding to the gate displacement field with $\mathcal{D}_c\approx0.1$~V/nm, is consistent with the experimental observations~\cite{han2024a}. By further increasing $\Delta$, Ising SOC term acts as a Haldane mass and $\Delta$ serves as the sublattice potential difference in the monolayer graphene. More precisely, in the SM phase, the low energy band consists of two valleys with the same spin. Therefore, the system is effectively spinless, the same as the Haldane model. If we consider the top and bottom layers as the A/B sublattices in the honeycomb lattice, the potential $\Delta$ represents the sublattice potential difference. Since the Ising-SOC has opposite signs for the two valleys, this term indeed maps to the Haldane mass\cite{han2023b}. Then, the QAH state emerges between the LAF state and LP state. This precisely realized the Haldane model for QAH~\cite{haldane1988}. And the LAF-$x$ state resembles the planar-AFM state in the Kane-Mele-Hubbard model~\cite{hohenadler2012}.

For experimental relevance, we limit the numerical calculation of the phase diagram with $\lambda$ value up to $2$~meV. However, for significantly larger, albeit spurious, $\lambda$ values, one could argue, based on the band edge alignment shown in Fig.~\ref{fig1}(b), that the LAF-$x$ state would undergo a crossover to the surface in-plane ferromagnetic state. Moreover, due to lack of inversion symmetry of Ising SOC, the phase diagram exhibits asymmetry with respect to $\Delta$. Specifically, the QAH state occupies a significant part in the phase diagram when $\Delta$ is less than zero, compared to its small occupation when $\Delta$ is greater than zero. Additionally, the critical value of $\lambda$ is lower when $\Delta<0$ than that when $\Delta>0$. This is consistent with the experimental observations, as the QAH effect has only been observed on one side of the displacement field for pentalayer graphene~\cite{han2023b}.

We further study the impact of Hund's coupling. The phase diagram in ($\Delta,J_\text{H}$) plane is depicted in Fig.~\ref{fig2}(a) for $\lambda=1$~meV. The phase boundary between LAF-$x$ state and SM state disperses linearly in $J_\text{H}$. To understand this, we noticed that the mean field decomposition of Eq.~(\ref{H_int}), for LAF state with either ordering directions, contains the mass term as $-J_{\text{H}}\mathbf{S}_{i,K}\cdot \mathbf{S}_{i,K'}\rightarrow-J_{\text{H}}(\mathbf{S}_{i,K}\cdot \langle \mathbf{S}_{i,K'}\rangle+\langle\mathbf{S}_{i,K}\rangle\cdot \mathbf{S}_{i,K'})$. Therefore, for $\Delta=0$, the gap of the LAF state is expected to increase approximately linearly with $J_{\text{H}}$. With the finite potential energy $\Delta$, we can naturally obtain a linear phase boundary. 
 
In particular, we discovered that the QAH state is suppressed as the strength of Hund's coupling $J_{\text{H}}$ increases. As we have discussed, the QAH state in this system can be modeled using the sign of the mass term of the four flavors, for example $(+,+,+,-)$ for $(K\uparrow, K\downarrow, K'\uparrow, K'\downarrow)$. It is noted that the amplitudes of mass term of these four flavors are generally unequal, leading to different spin polarizations in the two valleys, for example $|\langle\mathbf{S}_{K'} \rangle|\gg|\langle\mathbf{S}_K\rangle|\approx 0$. However, the Hund's coupling tends to favor a ferromagnetic order with similar spin densities at two valleys. In the above example, the spin density for the $K$ valley is much smaller than that for the $K'$ valley. Consequently, the Hund's coupling suppresses the QAH state as found in our numerical calculations.

Figs.~\ref{fig2}(b)-\ref{fig2}(e) show band structures of several typical phases. In the case of LAF-$x$, the colors represent the expectation value of spin polarization $\langle s_x\rangle$. For all other cases with spin $U(1)_s$ symmetry, the colors denote the layer polarization $\langle\varrho_z \rangle$. Explicitly, we see that the band gap for LAF-$x$ state is larger than that of LAF-$z$ state. Furthermore, the layer polarization of the QAH state is consistent with our analysis, where one flavor exhibits polarization opposite to the other three. This also indicates that the QAH state here has the Chern number $\mathcal{C}=5$. 
For the SM phase, it originated from the warping effect of the band structure with one electron pocket. Depending on the parameters, it also features either one annular hole pocket or three smaller hole pockets. Notably, we find that in the pentalayer system, even without SOC, the SM phase exhibits a non-vanishing anomalous Hall effect. Furthermore, based on orbital magnetization, a finite magnetic field with strength $B_z=1$~T can drive the system into a fully gapped QAH state, in agreement with experimental observations ~\cite{han2024a,supple}.

\begin{figure}[t]
\begin{center}
\includegraphics[width=3.4in, clip=true]{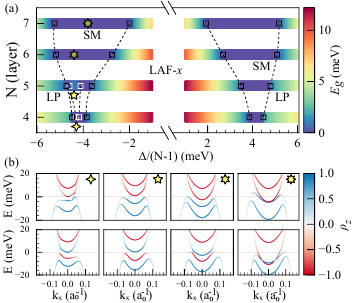}
\end{center} 
\caption{(a) Phase diagram with band gap of $N$ layer rhombohedral graphene vs $\Delta/(N-1)$ with $U=40$~\text{eV}$\cdot\mathcal{A}_u$, $V=-U/5$, $J_{\text{H}}=10$~\text{eV}$\cdot\mathcal{A}_u$, and $\lambda=1~\text{meV}$. The dashed lines are the phase boundaries for better illustration only. The black dashed lines represent the LAF-SM and LP-SM phase boundaries, while the white dashed region marks the QAH region. The QAH state for tetralayer only appears in a narrow region. (b) Hartree-Fock band structure at selected points in (a), with top (bottom) panel for $K$ ($K'$) valley. Stronger band warping effect in hexalayer and heptalayer drives the QAH state to be SM.}
\label{fig3}
\end{figure}

\emph{Layer dependence.} With a comprehensive understanding of the phase diagram for pentalayer graphene, we then turn our attention to layer dependence. Although systems with different thicknesses should be described by different interaction parameters $(U, V, J_{\text{H}})$, this would introduce complexity and uncertainty in exploring the layer-dependent phase diagram. Therefore, comparing with the experimental results for pentalayer graphene~\cite{han2024a,han2023b}, we fix the parameters for all the layers to be $\lambda=1$~meV, $J_{\text{H}}=10$~\text{eV}$\cdot\mathcal{A}_u$ and $U=40$~\text{eV}$\cdot\mathcal{A}_u$, $V=-U/5$. Fig.~\ref{fig3}(a) shows the numerical results ranging from tetralayer to heptalayer. 

For thickness greater than five, the QAH state becomes SM and is no longer insulating due to the strong warping effect as shown in Fig.~\ref{fig3}. In tetralayer graphene, the QAH state has a smaller gap compared to that in pentalayer for $\Delta<0$. Meanwhile, for $\Delta>0$, even though the QAH state now becomes SM, the indirect gap is much smaller in tetralayer. This explains well the experimental observations, where the anomalous Hall effect was found at zero magnetic field, but only became quantized in the presence of a small magnetic field~\cite{sha2023}. Remarkably, the critical field $\Delta_c$ for the LAF-$x$ state to SM transition is quantitatively consistent with experiments for various thicknesses\cite{han2024a,liu2024,han2023b,sha2023,zhou2024}. While for the second critical value of transition from SM to LP insulator, the disagreement may arise due to the fact that we may not simply relate $\Delta$ to the external field $\mathcal{D}$ without considering the screening effect~\cite{koshino2010}.

\begin{figure}[t]  
\begin{center}
\includegraphics[width=3.4in,clip=true]{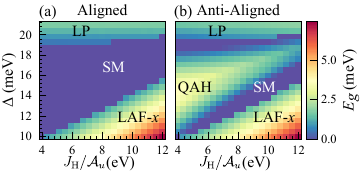}
\end{center} 
\caption{(a) \& (b) Hartree-Fock phase diagram of TMD-pentalayer graphene-TMD sandwich structure for the aligned ($\lambda^t=\lambda^b=1$~meV) and anti-aligned ($\lambda^t=-\lambda^b=1$~meV) cases, respectively. The QAH state is enlarged in the anti-aligned configuration. $U=40$~\text{eV}$\cdot\mathcal{A}_u$ and $V=-U/5$.}
\label{fig4}
\end{figure}

\emph{Sandwich configuration.} Finally, we consider the phase diagram of the sandwich configuration, namely TMD-graphene-TMD. Here the proximity Ising SOC would appear in both the top and bottom graphene layers with strengths $\lambda^t$ and $\lambda^b$, respectively. The Ising SOC in TMD is odd under $C_{2z}$ rotation. Thus, with two TMD materials neighboring the top and bottom layers, there are two configurations. One is the aligned case, where $\lambda^t$ and $\lambda^b$ have the same sign. The other is the anti-aligned case\cite{Island2019}, with one TMD being rotated $180^\circ$ relative to the other, and $\text{sgn}(\lambda^t)=-\text{sgn}(\lambda^{b})$. For simplicity, we set $|\lambda^t|=|\lambda^b|\equiv\lambda=1$~meV, and the Hamiltonian for these two cases is described by $\lambda\tau_z s_z$ and $\lambda\tau_zs_z\varrho_z$, respectively. Fig.~\ref{fig4} shows the numerical phase diagram for pentalayer graphene. For both cases, the phase diagram is symmetric with respect to the sign of $\Delta$. For the aligned case, since Ising SOC acts cooperatively on the two surfaces, the QAH state is suppressed. Unlike the one-side proximity case, which induces the term $\lambda\tau_z s_z (\varrho_0-\varrho_z)/2$, the anti-aligned case is described by $\lambda \tau_z s_z\varrho_z$. As a result, only the mass term from the one-side proximity case remains in the anti-aligned configuration, which enlarges the QAH state in the phase diagram. Therefore, such engineering of Ising SOC in RMG will stabilize the QAH effect with a full gap.

\emph{Discussion.} Our study provides a starting point for further theoretical and experimental investigation of correlated insulating phases in RMG with SOC. It also shows that in the presence of Ising SOC, the system exhibits an in-plane LAF order, which breaks the spin $U(1)_s$ symmetry. Therefore, the system should have gapless collective excitations, namely magnons. At finite temperature, the Berezinskii-Kosterlitz-Thouless phase transition is expected. Without SOC, the LAF state breaks the spin $SU(2)_s$ symmetry. It would be interesting to find the signature of the gapless magnon in future studies, for instance through spin transport. We notice, experimentally, that a $\mathcal{C}=3$ state was stabilized by a magnetic field near the metal-insulator transition in pentalayer graphene~\cite{han2024a}. Such a state cannot be simply obtained from current mean field theory. One possibility is the Hall crystal, which spontaneously breaks the translation symmetry~\cite{dong2023a,Tan2024,Dong2024a,Zeng2024,Sheng2024,Song2024,Soejima2024,Kwan2023,Guo2024}, another may be the Landau level. We leave these to future studies.

The interplay between interaction and SOC in proximitized RMG provides a versatile production line for the QAH effect with tunable Chern number. Remarkably, the rich choice of QAH states with different layer numbers, when proximitized to superconductivity, is promising to realize the exotic physics of chiral Majorana edge fermions and chiral Andreev edge states~\cite{qi2010b,wang2015c,lian2018b,wang2018,lian2019,Zhao2020a,uday2024,wang2024}.

\begin{acknowledgments}
\emph{Acknowledgment.} We acknowledge Guorui Chen and Yuanbo Zhang for helpful discussions. This work is supported by the Natural Science Foundation of China through Grants No.~12350404 and No.~12174066, the Innovation Program for Quantum Science and Technology through Grant No.~2021ZD0302600, the Science and Technology Commission of Shanghai Municipality under Grants No.~23JC1400600, No.~24LZ1400100, and No.~2019SHZDZX01.
\end{acknowledgments}

\end{document}


\title{Supplementary Material for ``Layer-Dependent Quantum Anomalous Hall Effect in Rhombohedral Graphene''}
\author{Zhaochen Liu}
\affiliation{State Key Laboratory of Surface Physics and Department of Physics, Fudan University, Shanghai 200433, China}
\affiliation{Shanghai Research Center for Quantum Sciences, Shanghai 201315, China}
\author{Jing Wang}
\thanks{wjingphys@fudan.edu.cn}
\affiliation{State Key Laboratory of Surface Physics and Department of Physics, Fudan University, Shanghai 200433, China}
\affiliation{Shanghai Research Center for Quantum Sciences, Shanghai 201315, China}
\affiliation{Institute for Nanoelectronic Devices and Quantum Computing, Zhangjiang Fudan International Innovation Center, Fudan University, Shanghai 200433, China}
\affiliation{Hefei National Laboratory, Hefei 230088, China}
	
\maketitle

\tableofcontents
	
\section{Single Particle Hamiltonian} 
Here, we present the Hamiltonian for ABC-stacking multilayer graphene. Taking tetralayer as example, in the basis of $(A_1,B_1,A_2,B_2,A_3,B_3,A_4,B_4)$, following the Refs.~\cite{jung2013a,ghazaryan2023}, 
the Hamiltonian near the $K$ and $K'$ point can be written as 
\begin{equation}
  H_{\tau,\text{tetralayer}}=\left(\begin{array}{cccc}
      h_0(1) & h_1 & h_2 & 0\\ 
      h^\dagger_1 & h_0(2) &h_1 & h_2 \\
      h^\dagger_2 & h^\dagger_1 & h_0(3) & h_1 \\
      0 & h^\dagger_2 & h^\dagger_1 & h_0(4) \\
    \end{array}
  \right),
  \label{eq:ham_ABC}
\end{equation}
where 
\begin{equation}
  h_{\tau,0}(i)=\left(\begin{array}{cc}
      \delta(2i-1) & v_0\pi^\dagger \\ 
      v_0 \pi  &\delta(2i)\\ 
  \end{array}
  \right),\quad 
  h_{\tau,1}=\left(\begin{array}{cc}
      v_4\pi^\dagger & v_3\pi \\ 
      \gamma_1  &v_4\pi^\dagger\\ 
  \end{array}
  \right),\quad 
  h_{\tau,2}=\left(\begin{array}{cc}
      0 & \gamma_2/2\\ 
      0 & 0\\ 
  \end{array}
  \right),
  \label{eq:ham_ABC_block}
\end{equation}
and $\pi=\tau k_x+ik_y$, $\tau=\pm 1$ for $K$ and $K'$ valley. The $\delta(i)=0$ for $i=1$ or $2N$ and equals to $\delta$ for others. And 
$v_i=\sqrt{3}a\gamma_i/2$, with $a=0.246$~nm as the lattice constant of graphene.
The parameters are chosen from Refs.~\cite{zhou2021,zibrov2018,ghazaryan2023} with 
\begin{equation}
  (\gamma_0,\gamma_1,\gamma_2,\gamma_3,\gamma_4,\delta)=(3.1,0.38,-0.015,-0.29,-0.141,0.0105)~\text{eV}
  \label{eq:para}
\end{equation}
\begin{figure}[htb]
  \begin{center}
    \includegraphics[width=0.3\textwidth]{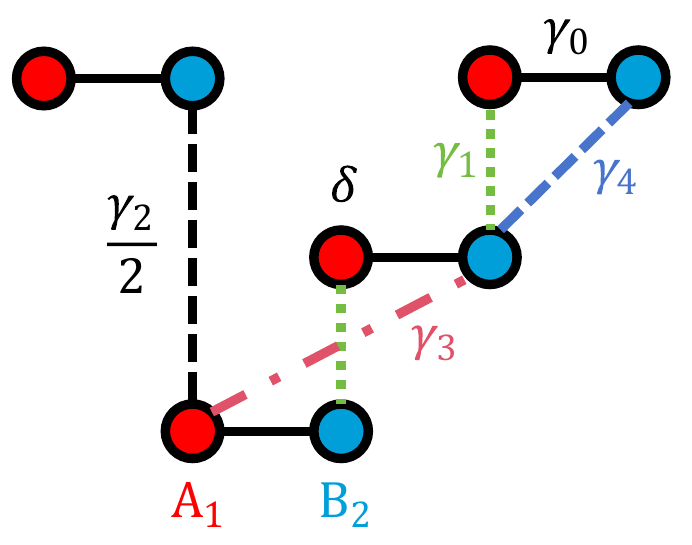}
  \end{center}
  \caption{Schematic for hopping and onsite terms of the tight-binding model.}
  \label{fig:hopping}
\end{figure}
In Fig.~\ref{fig:hopping}, we present the schematic of the hopping parameters of the tight-binding model.
The Hamiltonian for other thickness can be obtained by making an analogy with Eq.~(\ref{eq:ham_ABC}).
The staggered potential term is 
\begin{equation}
  H_{\tau,\text{tetralayer}}=\left(\begin{array}{cccc}
      \Delta_1 & 0 & 0 & 0\\ 
      0 & \Delta_2 & 0 & 0 \\
      0 & 0 & \Delta_3 & 0 \\
      0 & 0 & 0 & \Delta_4 \\
    \end{array}
  \right),
  \label{eq:potential}
\end{equation}
with $\Delta_i=\Delta \sigma_0(N-1-2(i-1))/(N-1)$ where $N$ is the thickness. The Ising-SOC term is 
\begin{equation}
  H^{\text{soc}}_{s,\tau,\text{tetralayer}}=\lambda\left(\begin{array}{cccc}
      0 & 0 & 0 & 0\\ 
      0 & 0 & 0 & 0 \\
      0 & 0 & 0 & 0 \\
      0 & 0 & 0 & s\tau \sigma_0\\
    \end{array}
  \right).
  \label{eq:soc}
\end{equation}

\subsection{Band structure warping effect}
As discussed in the main text, due to the strong band warping effect in 6L and 7L system, only the semi-metal phase exists between the LAF state and 
the LP state in the phase diagram. Therefore, if this warping effect can be suppressed~\cite{koshino2009a}, it may be possible to realize higher Chern number QAH states. In this section, 
we explore the origin of the band warping effect and discuss potential methods for its suppression.

By looking carefully at the Hamiltonian Eq.(\ref{eq:ham_ABC}), we notice when $\gamma_4=\delta=0$, the system has chiral symmetry. Namely, 
\begin{equation}
  C=I_{L}\otimes \sigma_z, C^\dagger H(\mathbf{k})C=-H(\mathbf{k})
\end{equation}
where $I_{n}$ is N dimensional identity operator. In this situation, if there is a band with energy $E(\mathbf{k})>0$, there will be a band with 
energy $-E(\mathbf{k})<0$. Thus, the band structure is symmetric with respect to $E=0$. The band warping effect is absent here. 

To further gain insights on the how does $\gamma_4$ and $\delta$ influence the band wrapping effect, we can project the Hamiltonian onto the low energy subspace
\cite{zhang2010a,koshino2009a,ghazaryan2023}.
Through the perturbation theory, the effective Hamiltonian is 
\begin{equation}
  H_{\text{eff}}=\frac{v^N_0k^N}{(-\gamma_1)^{N-1}}(\cos(N\varphi_k)\rho_x+\sin(N\varphi_k)\rho_y)-(\frac{2v_0v_4 k^2}{\gamma_1}+\frac{\delta v^2_0k^2}{\gamma_1})\rho_0+\cdots
  \label{eq:ham_eff}
\end{equation}
where $\cdots$ represents all other high order terms which may not qualitatively influence our conclusion.
The eigenvalue is
\begin{equation}
  E_{\pm}(k)=-(\frac{2v_0v_4 k^2}{\gamma_1}+\frac{\delta v^2_0k^2}{\gamma_1})\pm \frac{v^N_0k^N}{\gamma_1^{N-1}}
\end{equation}
as $-2v_4>\delta v_0$, the warping energy scale is determined by the valance band which is consistent with full band structure calculation.
Then, the energy scale of the warping effect is 
\begin{equation}
  E_{w}=E_{-}(k_c)-E_{-}(k=0) \quad \frac{dE_{-}}{dk}(k_c)=0
  \label{eq:warp_E}
\end{equation}
The finial result is quite complicated, therefore, for simplicity, consider $\delta=0$. We have $E_{w}\propto \gamma_1 (v_4/v_0)^{n/n-2}$. Ideally, one 
might expect that reducing the interlayer hopping $\gamma_1,\gamma_4$ could suppress the wrapping effect. However, experimentally, this cannot be done easily, 
as it requires increasing the interlayer distance. On the other hand, we can reduce the in-plane lattice constant through strain effects to enhance $\gamma_0$.

\begin{figure}[htb]
  \begin{center}
    \includegraphics[width=0.3\textwidth]{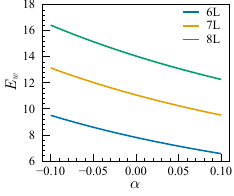}
  \end{center}
  \caption{Band warping energy for different thickness as function of effective strain parameter $\alpha$.}
  \label{fig:warping energy}
\end{figure}

In Fig.~\ref{fig:warping energy}, assuming the intralayer coupling is $(1+\alpha)\gamma_0$,
we present the dependence of $E_{w}$ on the phenomenological strain parameter $\alpha$ based on the full tight-binding model, . 
Clearly, our simple effective model calculation captures 
the main feature of $E_{w}$. The $E_{w}$ can only be reduced slightly when we increase $\gamma_1$ to a reasonable value. Therefore, we conclude that the strain effect 
cannot be used to efficiently suppress the warping effect.

\section{Self-Consistent Hartree-Fock Method}
Next, we will derive the Hartree-Fock equations. The interaction Hamiltonian can be written into two terms
\begin{equation}
  \begin{aligned}
  &H_{\text{U-V}}=\int d^2x \frac{U}{2}\sum_i n_i^2+ Vn_{i,K}n_{i,K'},\quad 
  H_{\text{J}}=-\int d^2x J_{\text{H}}\sum_i \mathbf{S}_{i,K}\cdot \mathbf{S}_{i,K'}
  \end{aligned}
  \label{eq:ham_int_2part}
\end{equation}
with the Fourier transformation $c_{i,\tau,s}(\mathbf{r})=\frac{1}{A^{1/2}}\sum_{\mathbf{k}} e^{i\mathbf{k}\cdot r}c_{i,\mathbf{k},\tau,s}$ and $A$ is area of the system.
We can obtain
\begin{equation}
  \begin{aligned}
    &H_{\text{U-V}}=\frac{U}{2A}\sum_{i,\mb{k},\mb{k'},\mb{q}}c^\dagger_{i,\mb{k+q},\alpha}c_{i,\mb{k},\alpha}c^\dagger_{i,\mb{k'-q},\beta}c_{i,\mb{k'},\beta}
         +\frac{V}{A}\sum_{i,\mb{k},\mb{k'},\mb{q},s,s'} c^\dagger_{i,\mb{k+q},K,s}c_{i,\mb{k},K,s} c^\dagger_{i,\mb{k'-q},K',s'}c_{i,\mb{k'},K',s'},
         \\ 
         &H_{\text{J}}=-\frac{J_{\text{H}}}{A}\sum_{i,\mb{k},\mb{k'},\mb{q},j,s,s',s'',s'''}c^\dagger_{i,\mb{k+q},K,s}c_{i,\mb{k},K,s'}
         c^\dagger_{i,\mb{k'-q},K',s''}c_{i,\mb{k'},K',s'''}\sigma^j_{s,s'}\sigma^j_{s'',s'''},
   \end{aligned}
  \label{eq:ham_fourier}
\end{equation}
where $\alpha,\beta=(\tau,s)$ represents the valley and spin indices. Within the Hartree-Fock method, assuming a translation symmetry preserving ansatz, we can obtain the 
Hartree term as 
\begin{equation}
  \begin{aligned}
  &h_{U,H}=\frac{U}{A}\sum_{i,\mb{k},\beta}\left[\sum_{k',\alpha}P^i_{\alpha,\alpha}(\mb{k'})\right]c^\dagger_{i,\mb{k},\beta}c_{i,\mb{k},\beta},\\
  &h_{V,H}=\frac{V}{A}\sum_{i,\mb{k},\tau,s}\left[\sum_{k',-\tau,s'}P^i_{-\tau,s';-\tau,s'}(\mb{k'})\right]c^\dagger_{i,\mb{k},\tau,s}c_{i,\mb{k},\tau,s},\\ 
  &h_{J,H}=-\frac{J_H}{A}\sum_{\mb{k},\tau}\left[\sum_{\mb{k'},s,s'}P^i_{\tau,s;\tau,s'}(\mb{k'})\sigma^j_{s,s'}\right]c^\dagger_{i,\mb{k},-\tau}\sigma^jc_{i,\mb{k},-\tau},
  \end{aligned}
  \label{eq:ABC_hartree}
\end{equation}
and the Fock term as 
\begin{equation}
  \begin{aligned}
    &h_{U,F}=-\frac{U}{A}\sum_{i,\mb{k},\alpha,\beta}\left[\sum_{\mb{k'}}P^i_{\alpha,\beta}(\mb{k'})\right]c^\dagger_{i,\mb{k},\beta}c_{i,\mb{k},\alpha},\\
    &h_{V,F}=-\frac{V}{A}\sum_{i,\mb{k},\tau,s,s'}\left[\sum_{\mb{k'}}P^i_{\tau,s;-\tau,s'}(\mb{k'})\right]c^\dagger_{i,\mb{k},-\tau,s'}c_{i,\mb{k},\tau,s},\\ 
    &h_{J,F}=\frac{J_{H}}{A}\sum_{\mb{k},\tau,s,s',s'',s'''} \left[\sum_{\mb{k'}} P^i_{\tau,s;-\tau,s'''}(\mb{k'})\right] c^\dagger_{i,\mb{k},-\tau,s''}c_{i,\mb{k},\tau,s'}
    \sigma^j_{s,s'}\sigma^j_{s'',s'''},
  \end{aligned}
  \label{eq:ABC_fock}
\end{equation}
where $P^i_{\alpha,\beta}(\mb{k})=\langle c^\dagger_{i,\mb{k},\alpha}c_{i,\mb{k},\beta}\rangle$. The expectation value is defined relative to the ground state of 
the Hartree-Fock Hamiltonian, $h_{HF}=h_0+h_{H}+h_{V}$. And the Hartree-Fock ground state energy is $E_G=\langle h_0+(1/2)(h_H+h_V)\rangle$.
As we can see, $h_{V,F}$ and $h_{V,J}$ is non-vanishing only after the system breaks the valley $U_v(1)$ symmetry.

The self-consistent Hartree-Fock equation is solved using the fix-point iteration method. 
Specifically, starting with an initial correlation matrix $P_n$, we construct the corresponding Hartree-Fock Hamiltonian $h_{n,HF}$ using Eq.~(\ref{eq:ABC_hartree}) and Eq.~(\ref{eq:ABC_fock}),
where $n$ denotes the iteration steps. We then obtain the updated correlation matrix $P_{n+1}$. However, it is important to note that the plain iteration method can be 
numerically unstable. To achieve stable solutions, we introduce a damping factor $\alpha$ and mix two matrices from the last two iterations. Thus, we use 
$\tilde{P}_{n+1}=(1-\alpha)P_{n}+\alpha P_{n+1}$ as the input for next iteration. In practice, we choose $\alpha=0.4$. If  
convergence issues arise, we can further reduce the value of $\alpha$, setting it to $0.1$ or even smaller.

To numerically solve above self-consistent equations, we discretize the momentum space into grids. 
In ensure the momentum grids satisfy the discrete symmetry of a single valley~\cite{koh2024}, we choose the momentum grid as:
\begin{equation}
  \mathbf{k}=n_1 \frac{\mathbf{g}_1}{N}+n_2 \frac{\mathbf{g}_2}{N},
  \label{eq:}
\end{equation}
where $\mathbf{g}_{1/2}=2\pi(1,\pm 1/\sqrt{3})/a_0$ is the reciprocal lattice of graphene and $a_0=0.246$nm is the lattice constant. $N$ is an integer number and $n_1,n_2$ range from 
$0$ to $N-1$. This momentum grid will fill the first Brillouin zone. Since we only need momenta near the $K$ and $K'$ points, we only consider momentum points near the 
$K/K'=(\pm 4\pi/3a_0,0)$ point with a cutoff $\Lambda$, i.e., only $|\mathbf{k}-\mathbf{K}|<\Lambda$. By taking $N=3N'$ where $N'$ is integer number, the momentum grid near 
$K$ and $K'$ valley coincides. Additionally, the system area is $A=N^2A_u=N^2\sqrt{3}a_0^2/2$. In practice, we take $\Lambda=0.16/a_0$ and $N=3\times 360$, resulting in a total 
total of approximately two thousand points inside the cutoff, which is sufficient for convergence.

\section{More Hartree-Fock Results}
\subsection{Phase diagram for bilayer and trilayer graphene}
In the main text, we studied the phase diagram for 4, 5, 6 and 7-layer rhombohedral graphene at charge neutrality. 
For completeness, we now provide the Hartree-Fock phase diagram for bilayer and trilayer graphene at charge neutrality as well. 
Before presenting the Hartree-Fock results, 
we note that bilayer and trilayer graphene exhibit a much smaller DOS compared to tetralayer and pentalayer systems. Therefore, 
at the charge neutrality with $\Delta=\lambda=0$, we expect these two systems to have significantly smaller band gaps. 
\begin{figure}[htb]
  \begin{center}
    \includegraphics[width=0.8\textwidth]{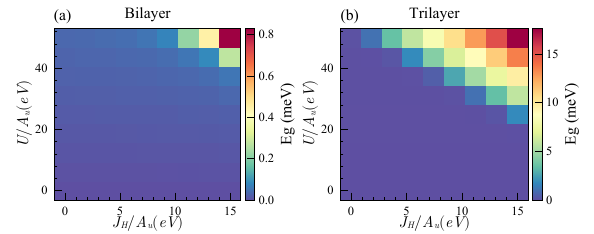}
  \end{center}
  \caption{(a,b) Band gap of the LAF state in bilayer and trilayer graphene without external displacement field and SOC, i.e., $\Delta=\lambda=0$. 
  And we set $V=0$.}
  \label{fig:LAF_23Lgap}
\end{figure}

Moreover, insulating states at charge neutrality have only been observed in free-standing samples~\cite{Bao2012,Bao2011}, indicating that nearby 
substrates can efficiently reduce the interaction strength. In Fig.~\ref{fig:LAF_23Lgap}, we present the band gaps of the LAF 
state at $\Delta=\lambda=0$ for bilayer and trilayer systems. By comparing these results with those for the pentalayer
(Fig.~\ref{fig:LAF_gap}(b,c)), we find that, in the vast majority of the phase space, 
these two systems remain gapless. A band gap only appears with sufficiently large interactions. 
As a result, we anticipate that the QAH state cannot be realized within the current mechanism, 
as achieving it would require tunning both the displacement field and maintaining charge neutrality, which necessitates a 
dual-gate configuration. This setup would inevitably reduce the interaction strength compared to the free-standing case, 
making the realization of the QAH state in the bilayer and trilayer graphene more challenging.

\subsection{Phase diagram for tetra- and pentalayer graphene}
\begin{figure}[b]
  \begin{center}
    \includegraphics[width=1\textwidth]{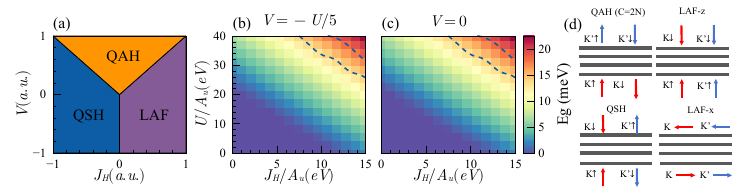}
  \end{center}
  \caption{(a) Schematic phase diagram in the $V-J_H$ plane when $\Delta=\lambda=0$.  (b,c) Band gap of the LAF state in pentalayer graphene without external displacement field and SOC, i.e., $\Delta=\lambda=0$. 
  (b)$V=-U/5$ and (c) $V=0$. Within the region enclosed by the dashed line, the band gap ranges from 13 to 17 meV. (d) Schematics of different symmetry-breaking states.}
  \label{fig:LAF_gap}
\end{figure}
As discussed in the main text, when $V=J_H=0$ and $\Delta=\lambda=0$, the QAH($C=2N$) state with mass term $\tau_z\varrho_z$, the LAF state with mass term $s_z\rho_z$ and 
the QSH state with mass term $\tau_z s_z\rho_z$ are degenerate. Analyzing the mass terms, we observe that the QAH($C=2N$) state has a non-vanishing valley imbalance, namely $n_{i,K}\neq n_{i,K'}$.
However, for the other two states, the electron densities of two valley are equal, $n_{i,K}=n_{i,K'}$.
Meanwhile, the LAF state corresponds to an inter-valley ferromagnetic state with $\mathbf{S}_{i,K}=\mathbf{S}_{i,K'}$ and the QSH state exhibits inter-valley antiferromagnetic order 
with $\mathbf{S}_{i,K}=-\mathbf{S}_{i,K'}$. Therefore, when $J_H=0$ and $V>0$, we expect the ground state to be the QAH state. When $J_H=0$ and $V<0$, the ground state 
becomes doubly degenerate, consisting of both the LAF and QSH states. On the other hand, when $V=0$, the ground state should be the LAF state when $J_H>0$ and the QSH state when $J_H<0$. The general phase diagram 
is shown in Fig.\ref{fig:LAF_gap}(a). Experimental observations have only found the trivial insulating state, leading us to conclude that $J_H>0$. 

Since our interaction term Eq.~(\ref{eq:ham_int_2part}) is a phenomenological Hamiltonian, it's not easy to obtain the corresponding coefficient
from first principles. From a renormalization group point of view, their values would also depend on the cutoff $\Lambda$. To determine these coefficients, we can compare our Hartree-Fock 
results with experimental observations. The experimental observations are as follows: Firstly, when $\Delta=\lambda=0$, the band gap is about 15~meV~\cite{liu2024}. Secondly, the QAH effect is
observed only for one sign of the displacement field, but not for the other~\cite{han2023b,sha2023}. Lastly, the maximum gap of the QAH effect is about 2~meV~\cite{han2023b}. 

Fig.~\ref{fig:LAF_gap} shows the band gap of the LAF state at $\Delta=\lambda=0$ for varying interaction strengths $U,V$ and $J_H$. In this scenario,
the LAF state is accurately described by the mass term $s_z\rho_z$, resulting in no valley imbalance and $n_{i,K}=n_{i,K'}$.
According to Eq.~(\ref{eq:ham_int_2part}), the $V$ term contributes only a constant energy, 
implying the band gap is independent of $V$. This is corroborated by the numerical results shown in Fig.~\ref{fig:LAF_gap}. 
The dashed line in Fig.~\ref{fig:LAF_gap} delineates the parameter region yielding a band gap $E_g\sim 15$meV. 
Additionally, we anticipate that the Coulomb repulsion energy $U$ should be stronger than Hund's coupling $J_H$. 
To further refine the interaction parameters, we computed the phase diagram and band gape for tetralayer and 
pentalayer with $\lambda=1$ meV for various combinations of $(U,V,J_H)$.

\begin{figure}[htb]
  \begin{center}
    \includegraphics[width=1.00\textwidth]{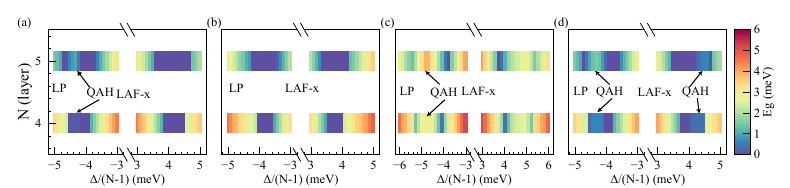}
  \end{center}
  \caption{Phase diagram and band gaps for pentalayer and tetralayer graphene under different potential energies $\Delta$. Here $\lambda=1$ meV. 
  (a) $(U,J_H)=(40,10)$~eV/$\mathcal{A}_u$, and $V=-U/5$. (b) $(U,J_H)=(35,12)$~eV/$\mathcal{A}_u$, and $V=-U/5$. In (c,d), $V=0$ with $U$ and $J_H$ the same as (a,b), respectively.}
  \label{fig:LAF_gap2}
\end{figure}
By comparing Fig.~\ref{fig:LAF_gap2}(a) and~\ref{fig:LAF_gap2}(c), we observe the $V<0$ suppressed the QAH state. 
Considering the mass term of the QAH state, for example $(+,+,+,-)$ for 
$(K\uparrow,K\downarrow,K'\uparrow,K'\downarrow)$, the electron density differs between the two valleys, i.e., $n_{i,K}\neq n_{i,K'}$. 
In contrast, there is no valley polarization for the 
LP state and LAF states. Therefore, when $V<0$, as indicated by Eq.~(\ref{eq:ham_int_2part}), the QAH state is suppressed. Additionally, as discussed in the main text, an increased Hund's coupling 
strength $J_H$ also suppresses the QAH state. By comparing the phase diagrams in Fig.~\ref{fig:LAF_gap2}, and without further fine-tunning the parameters, 
we propose that $(U,J_H)=(40,10)$~eV/$\mathcal{A}_u$ and $V=-U/5$ are suitable to describe the correlated physics in pentalayer and tetralayer graphene. 

\subsection{Phase diagram of the semi-metal state}
In this section, we present a detailed analysis of the semi-metal state that exists between the LAF and LP states. 
Experimentally, in the absence of Ising SOC, the 
semi-metal state has been observed in both tetralayer and pentalayer systems~\cite{han2024a,liu2024}. More interestingly, 
under a finite out-of-plane magnetic field, a Chern number $C=5$ QAH state emerges from 
the semi-metal state in the pentalayer system~\cite{han2024a}. Therefore, it is crucial to gain a better understanding of the semi-metal 
phase and the roles of Ising SOC and the magnetic field.

As we have shown, without Ising SOC ($\lambda=0$) and an external displacement field ($\Delta=0$), the ground state is the
LAF state. There are two distinct LAF states corresponding to the mass terms $(+,-,+,-)$ and $(-,+,-,+)$ for 
$(K\uparrow,K\downarrow,K'\uparrow,K'\downarrow)$. Taking the $(+,-,+,-)$ as an example, a finite external displacement field 
with $\Delta<0$ will decrease the band gap of ($K\uparrow,K'\uparrow$) and increase the band gap of ($K\downarrow,K'\downarrow$). 
Thus, the system effectively becomes spinless.
There will be a critical field $\Delta_c$ where the band gap of spin up fermions vanishes. Precisely at this critical field, 
one may view the Hartree-Fock band structure for spin up fermions as "non-interacting" band structure. Based on 
this Hartree-Fock band structure, an additional secondary mass term may further lower the energy. 
Moreover, since the DOS has been reduce by half and part of the interaction effect is already considered in the 
original mass term, we expect this secondary mass term to be much weaker compared
to the $\Delta=0$ case.

\begin{figure}[htbp]
  \begin{center}
    \includegraphics[width=0.95\textwidth]{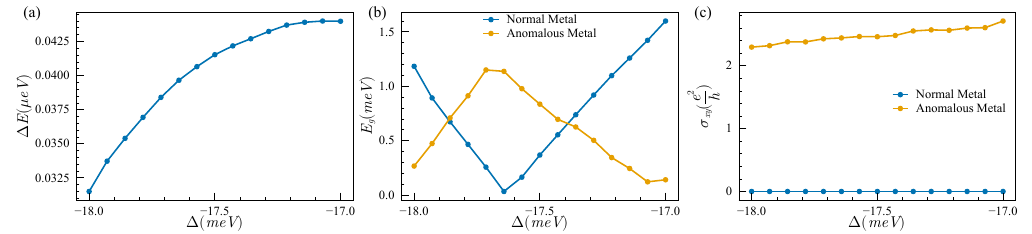}
  \end{center}
  \caption{Hartree-Fock results for the pentalayer system without Ising SOC in the semi-metal region.
  (a) Energy difference per unit cell between the normal semi-metal state and the anomalous semi-metal state. 
  (b) Band at the $\mathbf{k}=0$ point in the Hartree-Fock band structure. (c) Hall conductance of the anomalous semi-metal phase. The 
  parameters are $\Delta=-17.63$ meV, $U= 40~\text{eV}\cdot\mathcal{A}_u$, $V=-U/5$, and $J_H=10~\text{eV}\cdot\mathcal{A}_u$.}
  \label{fig:SM_Gap}
\end{figure}

If we denote the additional mass term as $(\delta m_{K,\uparrow},\delta m_{K',\uparrow})$, there are two possibilities.
When the two terms are equal, the system is in a normal semi-metal state. If the two masses have opposite signs, the system 
represents a semi-metal state with an anomalous Hall effect. There are two degenerate anomalous semi-metal phases with opposite 
Hall conductance.
In Fig.~\ref{fig:SM_Gap}, we present the Hartree-Fock ground state 
energy, direct band gap, Hall conductance, and orbital magnetization for two different semi-metal states as a function of the interlayer potential $\Delta$ 
for the pentalayer system. As we can see, the anomalous metal state has lower energy.
Interestingly, in the pentalayer case, the anomalous metal state has much lower energy than the normal semi-metal. However, for the tetralayer case,
they have almost the same energy. This difference can be attributed to the fact that tetralayer system has a lower density of states compared 
to the pentalayer system.

\begin{figure}[b]
  \begin{center}
    \includegraphics[width=0.95\textwidth]{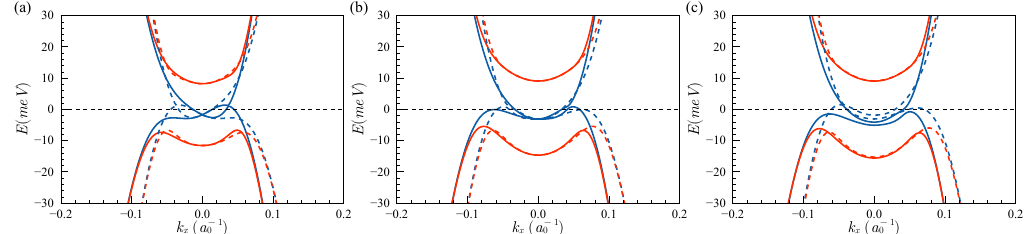}
  \end{center}
  \caption{Hartree-Fock band structure for the system without Ising SOC in the semi-metal phase at the critical $\Delta_c$. Blue and red lines represents the 
  spin up and down states. while the solid and dashed lines correspond to the $K$ and $K'$ valleys.
  (a) Band structure of the tetralayer system. 
  (b) Band structure of the normal semi-metal phase in the pentalayer system. (c) Band structure of the anomalous semi-metal state in the pentalayer system. The 
  parameters are $U= 40~\text{eV}\cdot\mathcal{A}_u$, $V=-U/5$, and $J_H=10~\text{eV}\cdot\mathcal{A}_u$ with $\Delta_c=-17.63$ meV for the pentalayer and 
  $\Delta_c=-13.05$ meV for the tetralayer.}
  \label{fig:SM_Band}
\end{figure}
Fig.~\ref{fig:SM_Band} shows the Hartree-Fock band structures at the critical potential $\Delta_c$. In this context, the normal semi-metal 
state has a gapless band structure, while the anomalous metal state has a finite direct band gap (with a vanishing indirect band gap).
It is this band gap that lowers the energy. Moreover, for the anomalous metal state,
the lack of particle-hole symmetry in the system causes our Hartree-Fock calculation to generate a valley-dependent potential term 
$\delta \mu\rho_0\tau_z$. The impact of this term is illustrated in Fig.~\ref{fig:SM_Band}(c).

\begin{figure}[htb]
  \begin{center}
    \includegraphics[width=0.85\textwidth]{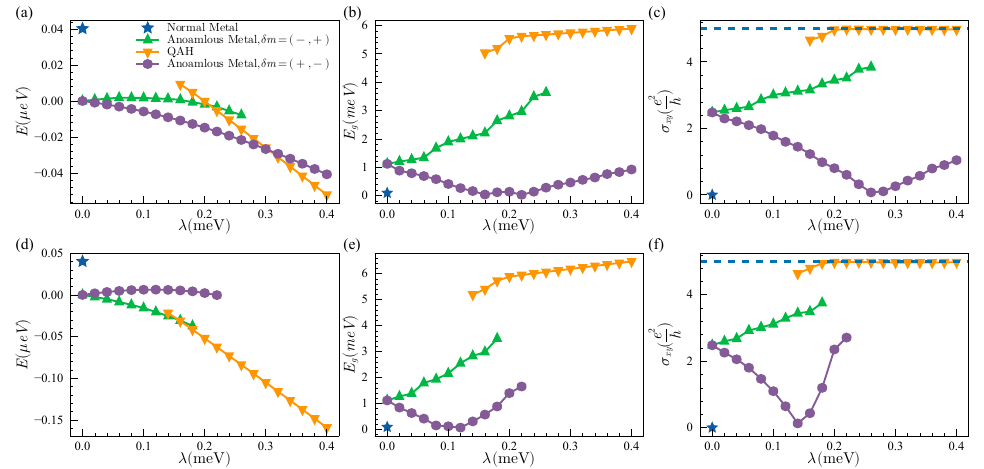}
  \end{center}
  \caption{Hartree-Fock results for the pentalayer system as function of SOC strength in the semi-metal region at the critical $\Delta_c$.
  The top panel shows the results for one-sided Ising SOC proximity system, and the bottom panel represents the results for an anti-aligned two-sided configuration.
  (a,d) The Hartree-Fock ground state energy per unit cell for different states.
  (b,e) Band gap at the $\mathbf{k}=0$ point of the Hartree-Fock band structure. (c,f) Absolute value of Hall conductance for the anomalous semi-metal phase. The 
  parameters are $\Delta=-17.63$ meV, $U= 40~\text{eV}\cdot\mathcal{A}_u$, $V=-U/5$, and $J_H=10~\text{eV}\cdot\mathcal{A}_u$.}
  \label{fig:SM_SOC}
\end{figure}

Next, we turn to the system with one side coupled wtih a monolayer TMD. For simplicity, we only consider the case at the critical $\Delta_c$.
For the normal semi-metal phase, there is a spinless time-reversal symmetry, $\mathcal{T}=\tau_x K$, where $K$ is the
complex conjugate operator. Since the Ising-SOC, $\lambda (\rho_0-\rho_z)\tau_z/2$, has opposite signs for the two valley, it breaks this 
spinless time-reversal symmetry. We may expect the anomalous semi-metal state with a similar mass term to Ising SOC to be the ground state.
This is indeed the case when the SOC strength is strong enough. However, at weak SOC strength, this may not be the case as
the valley dependent-chemical potential term in a one-sided SOC proximity system may complicate the result.
In Fig.~\ref{fig:SM_SOC} (a-b), we present the Hartree-Fock results for the system at the critical $\Delta_c$. For the pentalayer system, the anomalous 
semi-metal phase with a mass term opposite to the SOC term has lower energy when $\lambda<\lambda_c\sim 0.3\text{meV}$. 
When $\lambda>\lambda_c$, the Ising SOC will select the state with the mass term sign as $(-,+)$. Further increasing the SOC will 
enlarge the band gap and eventually lead to a fully gapped QAH state. 
For the tetralayer case, the energy difference between different states is very small. 
This originates from relatively weak density of states compared to the pentalayer case. Furthermore, 
we also present the results for anti-aligned sandwich 
structure in Fig.~\ref{fig:SM_SOC} (e-f). Here, the corresponding term in the Hamiltonian is simply $-\lambda \rho_z\tau_z$. In this situation, 
the SOC term only contains the mass term, thus it will only select the anomalous semi-metal state with the same sign of the mass term. This is consistent 
with our numerical calculations.

\section{Orbital magnetization and Role of magnetic field to QAH state}
\begin{figure}[htb]
  \begin{center}
    \includegraphics[width=0.95\textwidth]{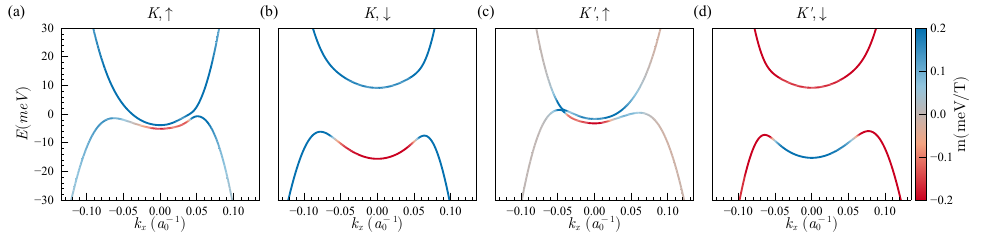}
  \end{center}
  \caption{Hartree-Fock band structures and orbital magnetic moment for the pentalayer system without Ising-SOC in the semi-metal region. 
  The parameters are the same as Fig.~\ref{fig:SM_Band}.}
  \label{fig:SM_Band_OrbitalM}  
\end{figure}

In the previous section, we showed that for the pentalayer system, the anomalous semi-metal phase has lower energy than the normal semi-metal state, 
and that Ising SOC can stabilize a gapped QAH state. Experimentally, an out-of-plane magnetic field can also induce a QAH state~\cite{han2024a}. 
In this section, we discuss the coupling between 
the magnetic field and the anomalous metal state. The magnetic field can couple to the Bloch band through orbital magnetic moment~\cite{THONHAUSER2011,Xiao2010,Xiao2007}.
The orbital magnetic moment of the Bloch electron is given by:
\begin{equation}
  m_{n}(\mathbf{k})=-i\frac{e}{2\hbar}\epsilon_{\mu,\nu}\langle \partial_{k_\mu} u_{n,\mathbf{k}}|h(\mathbf{k})-E_{n,\mathbf{k}}|
  \partial_{k_\mu} u_{n,\mathbf{k}}\rangle
\end{equation}
by noticing 
\begin{equation}
  \langle u_{m}(\mathbf{k})|\partial_{k_\mu} u_{n}(\mathbf{k})\rangle= -\frac{\langle u_{m}(\mathbf{k})|\partial_{k_\mu} h(\mathbf{k})|u_{n}(\mathbf{k})\rangle}
  {E_m(\mathbf{k})-E_n(\mathbf{k})}
\end{equation}
with $n\neq m$, we obtain 
\begin{equation}
  m_{n}(\mathbf{k})=\frac{e}{\hbar}\text{Im}\sum_{m\neq n}\frac{\langle u_{n,\mathbf{k}}|\partial_{k_\mu} h(\mathbf{k})|u_{m,\mathbf{k}}\rangle\langle u_{m,\mathbf{k}}|\partial_{k_\mu} h(\mathbf{k})|u_{n,\mathbf{k}}\rangle}{E_m(\mathbf{k})-E_n(\mathbf{k})}
  \label{eq:magnetic_moment}
\end{equation}
while for the total orbital magnetization, one need further account the Berry curvature correction to the density of states, which leads to 
\begin{equation}
  M_n=\int \frac{d^2k}{(2\pi)^2} f(E_{n}(\mathbf{k})-\mu)(m_{n}(\mathbf{k})+\frac{e}{\hbar}(\mu-E_{n}(\mathbf{k}))\Omega_n(\mathbf{k}))
\end{equation}
where $\Omega_{n}(\mathbf{k})=i\epsilon_{\mu,\nu}\langle \partial_{k_\mu}u_{n}(\mathbf{k})| \partial_{k_\nu}u_{n}(\mathbf{k})\rangle$ is the Berry curvature,
$E_{n}(\mathbf{k})$ is energy for n-th band with wave function $|u_{n}(\mathbf{k})\rangle$, and $f$ is Fermi-Dirac distribution. 
Fig.~\ref{fig:SM_Band_OrbitalM} shows the distribution of magnetic moment of the anomalous semi-metal state of the pentalayer system. Notably, at the band 
edge $\mathbf{k}\rightarrow 0$, the conduction and valance bands have opposite magnetic moments. For the two band mode, 
from Eq.~(\ref{eq:magnetic_moment}), the magnetic moments of conduction and valance bands are always equal, highlighting the importance of multi-band effects
in our case. Additionally, for spin up bands, the magnetic moments at the band edge of two valley are equal to each other, with a strength $m\sim 0.2 \text{meV/T}$.
Since the magnetic field modifies the band structure as $E_{n}(\mathbf{k})-m_{n}(\mathbf{k})B_z$~\cite{THONHAUSER2011}, it increases the band gap 
in both valleys. The band edges are polarized into opposite layers in the two valleys, meaning the effect of magnetic field can be phenomenologically 
described by $m B \rho_z\tau_z$, the same as two-sided anti-aligned structure. Comparing these findings with the Hartree-Fock 
results in Fig.~\ref{fig:SM_SOC}, we estimate the critical field to be approximately $B_c\sim 1\text{T}$ which aligns with 
experimental observations~\cite{han2024a}. 

\section{The role of Ising SOC to symmetry breaking states}
\subsection{One-sided RMG-TMD system}
As discussed in the main text and previous sections, the ground state in the absence of an external displacement field 
is the layer antiferromagnetic (LAF) state. However, when one side of the RMG is coupled to a TMD, a $\lambda (\rho_0-\rho_z)\tau_z s_z/2$ term is introduced,
breaking the inversion symmetry and opening band gaps. This term also induces a valley-spin-dependent chemical potential, leading to
a non-trivial band alignment. Based on the band alignment, we propose that the ground state could be either an in-plane LAF state or an inter-valley 
coherent (IVC) state. In this section, we present further results and discussions on the phase diagram in the presence of the Ising-SOC.

The band alignment argument could be understood in terms of excitonic instabilities~\cite{Jerome1967}. In the large $\lambda$ limit, as shown in Fig.1(b) of the main 
text, the low energy physics consist of conduction bands of the $(K,\uparrow,K'\downarrow)$ valleys, as well as the valance bands of the 
$(K,\downarrow,K'\uparrow)$ valleys, all of which are polarized to the top layer. This results two possible electron-hole pairings, one is intra-valley pairing
between $(K,\uparrow)$ and $(K,\downarrow)$, and $(K',\downarrow)$ and $(K',\uparrow)$, and another is inter-valley pairing between $(K,\uparrow)$ and $(K',\uparrow)$, and 
$(K',\downarrow)$ and $(K,\uparrow)$. In the first case, the spins align in the in-plane direction. 
Furthermore, due to the inter-valley ferromagnetic Hund's coupling, this corresponds to an in-plane surface ferromagnetic order with 
order parameter $s_{x/y}(\rho_0+\rho_z)/2$. In the small $\lambda$ limit, this state undergoes a crossover to an in-plane layer antiferromagnetic state with the order parameter $s_{x/y}\rho_z$.

While the second case is a inter-valley coherent state. Moreover, due to the spin Hund's couping can be related to the IVC Hund's coupling as shown in 
Eq.~(\ref{eq:valleyHundCoupling}). Thus, for the IVC order, the energy of IVC Hund's coupling is 
\begin{equation}
  E_{J}=-\sum_i 2J_H (\langle c^\dagger_{i,K,\uparrow}c_{i,K',\uparrow} \rangle-\langle c^\dagger_{i,K,\downarrow}c_{i,K',\downarrow} \rangle)
  (\langle c^\dagger_{i,K',\uparrow}c_{i,K,\uparrow} \rangle-\langle c^\dagger_{i,K',\downarrow}c_{i,K,\downarrow} \rangle)
\end{equation}
for $\langle c^\dagger_{i,K,\uparrow}c_{i,K',\uparrow} \rangle=\langle c^\dagger_{i,K,\downarrow}c_{i,K',\downarrow} \rangle e^{i \alpha}$, the energy is lowest for 
$\alpha=\pi$. Therefore, we expect the IVC order parameter to be $\tau_{x/y}s_z(\rho_0+\rho_z)/2$. This state can be stabilized in the Hartree-Fock 
calculations. Whether the ground state is the LAF-x or IVC state should be determined by the self-consistent Hartree-Fock calculation. 
To further characterize the IVC state, we define the order parameter as the norm of the off-diagonal part of the density matrix 
\begin{equation}
  O_{\text{IVC}}=\frac{1}{N_k}\sum_{\mathbf{k}}||P_{K,K'}||
\end{equation}
where $N_k$ is number of momentum points within the cutoff and  $||A||=\sqrt{\sum_{i,j}|A_{i,j}|^2}$.

\begin{figure}[htb]
  \begin{center}
    \includegraphics[width=0.9\textwidth]{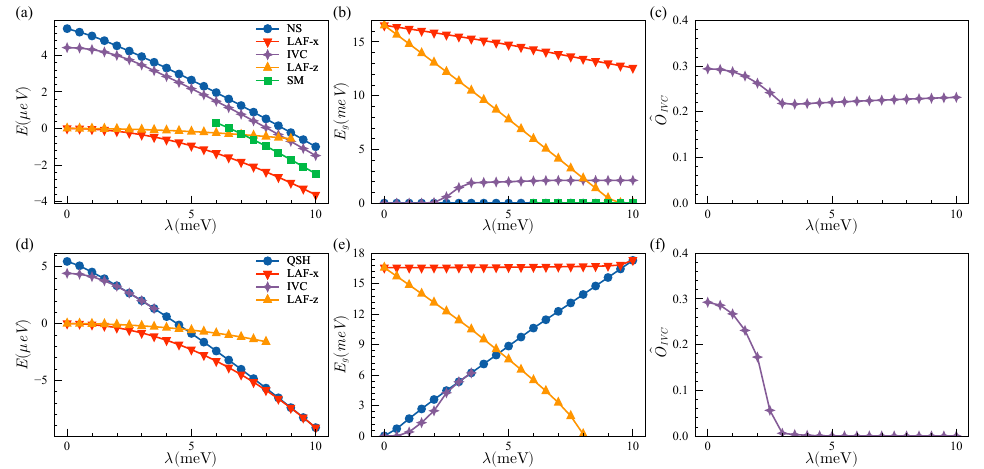}
  \end{center}
  \caption{Hartree-Fock results for the pentalayer system as a function of SOC strength without external displacement field. Results for the tetralayer 
  system are similar,thus not shown here.
  The top panel shows the results for one-sided Ising SOC proximity system, and the bottom panel shows the anti-aligned two-sided configuration.
  The "NS" refers to the non-symmetry breaking normal semi-metal state.
  (a,d) Hartree-Fock ground state energy per unit cell for different states, with the energy of the LAF state at $\lambda=0$ set as the reference.
  (b,e) The indirect band gap of the Hartree-Fock band structure. (c,f) The IVC state order parameter. The 
  other parameters are $U= 40~\text{eV}\cdot\mathcal{A}_u$, $V=-U/5$, and $J_H=10~\text{eV}\cdot\mathcal{A}_u$.}
  \label{fig:D0_SOC}
\end{figure}

In the top panel of Fig.~\ref{fig:D0_SOC}, we present the self-consistent Hartree-Fock results for various states as a function of the Ising SOC strength in 
the absence of an external displacement field for the pentalayer system. The IVC state exhibits much higher energy compared to the LAF state and only  
develops a band gap at a sufficient large SOC.
The band gap of the LAF-z state is suppressed by the SOC, while the gap of the LAF-x state decreases only slightly as the SOC strength increases. Furthermore, 
the SOC significantly lowers the energy of the LAF-z state relative to the LAF-x state. Interestingly, in the large $\lambda$ limit, we identify another semi-metal state, where 
the band in the $K'$ valley is gapped, with the $(K,\uparrow)$ valley hole doped and the $(K',\downarrow)$ valley electron doped.

\subsection{Two-sided system}
Next, we consider the sandwich configuration, specifically the TMD-RMG-TMD system, 
focusing on the anti-aligned setup with $\lambda_t=-\lambda_b=\lambda$. Unlike 
the one-sided coupled system, the Ising SOC in this case introduces only a mass term $\lambda \tau_z  s_z\rho_z$, 
contributing to a band gap without a valley-spin dependent chemical potential. As a result, the band alignment argument from the previous section no longer applies. 
Interestingly, the Ising SOC can be interpreted not only as a valley-spin-dependent mass term, but also as a valley-layer-dependent out-of-plane spin Zeeman term.
For the LAF state,
the impact of SOC can by analyzed using a phenomenological Ginzburg-Landau free energy approach. Assuming the spin densities of the top and bottom layer satisfies 
$\mathbf{s}_{\tau,t}=-\mathbf{s}_{\tau,b}$, the Ginzburg-Landau free energy density is 
\begin{equation}
  f=-J\mathbf{s}_{K,t}\cdot \mathbf{s}_{K',t}-J\mathbf{s}_{K,b}\cdot \mathbf{s}_{K',b}+\lambda s_{z,K,t}-\lambda s_{z,K',t}
  -\lambda s_{z,K,b}+\lambda s_{z,K',b}
\end{equation} 
Clearly, the ground state configuration is a valley-canted layer antiferromagnetic state, characterized by
$\mathbf{s}_{K,t}=|\mathbf{s}|(\cos{\theta},0,-\sin{\theta})$ and 
$\mathbf{s}_{K',t}=|\mathbf{s}|(\cos{\theta},0,\sin{\theta})$. The $\theta$ angle is determined by minimizing the free energy, 
yielding $\theta=\arctan{\lambda/\sqrt{4J^2 s^2-\lambda^2}}$. Although the spin density within each valley is canted, the total spin density
, denoted as $\mathbf{s}_i=\mathbf{s}_{K,i}+\mathbf{s}_{K',i}$, remains aligned in the in-plane direction. Thus, we still refer to this state as 
LAF-x state.
When the SOC exceeds the critical value, $\lambda>\lambda_c=\sqrt{4J^2 s^2-\lambda^2}$, the spins become completely out-of plane, 
corresponding to the QSH state. Conversely, in the IVC state, 
the two-sided Ising SOC term serve as a pair-breaking field, effectively suppressing the IVC state.

The Hartree-Fock results for the anti-aligned sandwich configuration are shown in bottom panel of Fig.~\ref{fig:D0_SOC}. The band gap of the LAF-x remains almost 
unchanged as $\lambda$ increases, and this state eventually crossover to the QSH state. Meanwhile, the IVC state is rapidly suppressed by the 
Ising SOC, consistent with the above analysis.

\section{Transition From Insulator to Semi-Metal}
As discussed in the main text and previous section, when the inter-layer potential $\Delta$ is below a critical value and 
Ising SOC is absent, the ground state is the LAF-x state. At the critical $\Delta_c$, a first-order phase transition occurs from the gapped 
LAF-x state to a semi-metal state. In this section, we provide a detailed discussion of the mechanism underlying this phase transition.

\begin{figure}[htb]
  \begin{center}
    \includegraphics[width=1.0\textwidth]{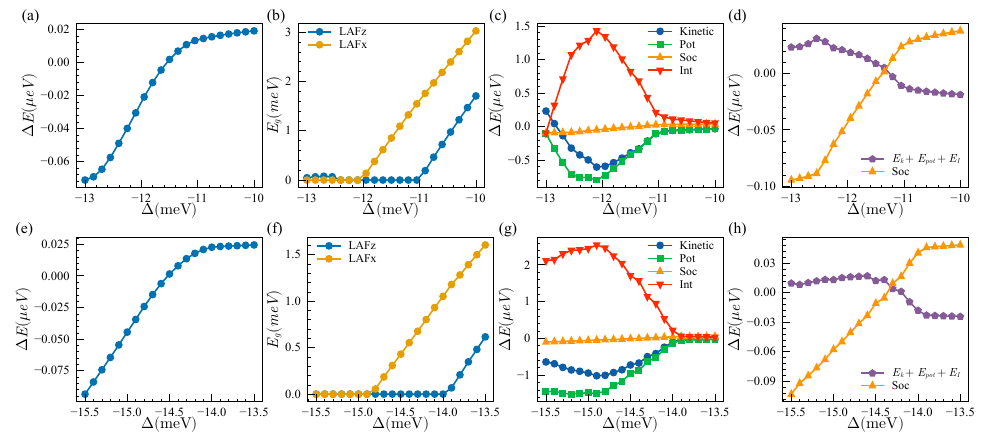}
  \end{center}
  \caption{Hartree-Fock results for the LAF-z and LAF-x states for the tetralayer and pentalayer systems as a function of the external displacement field 
  with finite SOC. The top panel shows results for the tetralayer, while the bottom panel represents the results for the pentalayer.
  (a,e) Energy difference per unit cell between the LAF-z state and LAF-x states.
  (b,f) The indirect band gap. (c,g) Energy difference per unit cell between the LAF-z and the LAF-x states for different components.
  (d,h) Energy difference per unit cell between the LAF-z and LAF-x states, separated into SOC part and non-SOC contributions. 
  The SOC part is responsible for the LAF-z state having a lower energy than the LAF-x state in the gapless region.
  The SOC strength is $\lambda=1\text{meV}$, and other parameters are $U= 40~\text{eV}\cdot\mathcal{A}_u$, $V=-U/5$, and $J_H=10~\text{eV}\cdot\mathcal{A}_u$.}
  \label{fig:InsulatorToSM_wSOC}
\end{figure}

Fig.~\ref{fig:InsulatorToSM_wSOC} presents the Hartree-Fock results for pentalayer and tetralayer system with one-sided SOC as a function of the interlayer layer 
potential $\Delta$. Interestingly, when the $\Delta$ is small, both the LAF-z and LAF-x states are gapped, with the LAF-x state having lower energy 
as we suggested. Additionally, the LAF-z state exhibits a smaller energy gap than the LAF-x state. Consequently, as $\Delta$ increases, the LAF-z state 
first transitions to a gapless semi-metal state. As discussed in the previous section, the Ising SOC breaks the spinless time-reversal symmetry in the low energy subspace, 
leading to an anomalous Hall effect in the semi-metal state. Furthermore, a first order phase transition happens from the gapped LAF-x state to the semi-metal state.
When we decompose the ground state energy into different components--kinetic, interlayer potential energy, Ising SOC energy, and interaction energy--we find the 
LAF-z state has lower kinetic and potential energy but higher interaction energy compared to the LAF-x state. The energy gain from the kinetic and potential 
terms nearly compensates for the energy loss from interaction term, with the energy gain from SOC term driving the phase transition.

Motivated by the numerical observations above, we aim to gain a deeper understanding of this phase transition.
To do so, we can treat the Ising SOC as a perturbation and analyze its impact on the ground state energy of the LAF state in the absence of Ising SOC.
To the second order in SOC strength, $\lambda$, the corresponding Feynman diagram for LAF state is shown in Fig.~\ref{fig:feyn_semi}. Where the solid line 
represents the Green's function of the Hartree-Fock hamiltonian, and the SOC vertex is dressed with interactions. We consider only the one-sided SOC-coupled case,
with Ising SOC as $\lambda \tau_z s_z (\rho_0-\rho_z)/2$.
\begin{figure}[htb]
  \begin{tikzpicture}[baseline=(current  bounding  box.center)]
    
    \begin{scope}[yshift=0mm,xshift=0mm,
      line width=1pt]
      \path[name path=C]  (1/3,1/3) -- (0,0) -- (1/3,-1/3);  
      \path[name path=X]  (1/3,-1/3)  (1/3,1/3);  
      \tikzfillbetween[of=C and X]{gray!90};
      \begin{feynman}
        \vertex (x) at ( 0, 0) ;
        \vertex (x1) at ( 1/3, 1/3) ;
        \vertex (x2) at ( 1/3, -1/3) ;
        \vertex (y) at (1.5, 0) ;
        \vertex (y1) at (2,0, 0) ; 
        \vertex[left= 0.6cm of x] (p);
        \diagram*{ 
            (p) --[scalar] (x), 
            (x) -- (x1);
            (x) -- (x2);
            (x1) --[fermion,out=45, in=120] (y),
            (y) --[fermion,out=-120, in=-45] (x2), 
            (y) --[scalar] (y1), 
        };
      \end{feynman}
      \node[below=10mm,xshift=10mm]    {(a)};
    \end{scope}

    \begin{scope}[yshift=0mm,xshift=30mm,
      line width=1pt,scale=0.6]
      \path[name path=C]  (1/3,1/3) -- (0,0) -- (1/3,-1/3);  
      \path[name path=X]  (1/3,-1/3)  (1/3,1/3);  
      \tikzfillbetween[of=C and X]{gray!90};
      \begin{feynman}
          \vertex (x) at ( 0, 0) ;
          \vertex[] (a) at (1, 1) ;
          \vertex[] (b) at (1,-1); 
          \vertex[left= 0.6cm of x] (p);
          \diagram*{ 
              (p) --[scalar] (x), 
              (x) --[fermion] (a),
              (b) --[fermion] (x), 
          };
      \end{feynman}
      \node[xshift=10mm]    {\large =};
    \end{scope}
    \begin{scope}[yshift=0mm,xshift=50mm,
      line width=1pt,scale=0.6] 
      \begin{feynman}
          \vertex (x) at ( 0, 0) ;
          \vertex[] (a) at (1, 1) ;
          \vertex[] (b) at (1,-1); 
          \vertex[left= 0.6cm of x] (p);
          \diagram*{ 
              (p) --[scalar] (x), 
              (x) --[fermion] (a),
              (b) --[fermion] (x), 
          };
      \end{feynman}
      \node[xshift=10mm]    {\large +};
    \end{scope}
    \begin{scope}[yshift=0mm,xshift=70mm,
      line width=1pt,scale=0.6]
      \path[name path=C]  (1/3,1/3) -- (0,0) -- (1/3,-1/3);  
      \path[name path=X]  (1/3,-1/3)  (1/3,1/3);  
      \tikzfillbetween[of=C and X]{gray!90};
      \begin{feynman}
          \vertex (x) at ( 0, 0) ;
          \vertex[] (a) at (1, 1) ;
          \vertex[] (b) at (1,-1);
          \vertex[right=1cm of a] (a1);
          \vertex[right=1cm of b] (b1);
          \vertex[left= 0.6cm of x] (p);
          \diagram*{ 
              (p) --[scalar] (x), 
              (x) --[fermion] (a),
              (a) --[fermion] (a1),
              (b) --[fermion] (x), 
              (b1) --[fermion] (b),
              (a) --[photon] (b),
          };
      \end{feynman}
      \node[xshift=-10mm]    {\large +};
    \end{scope}

    \begin{scope}[yshift=0mm,xshift=100mm,
      line width=1pt,scale=0.6]
      \path[name path=C]  (1/3,1/3) -- (0,0) -- (1/3,-1/3);  
      \path[name path=X]  (1/3,-1/3)  (1/3,1/3);  
      \tikzfillbetween[of=C and X]{gray!90};
      \begin{feynman}
          \vertex (x) at ( 0, 0) ;
          \vertex (x1) at ( 1/3, 1/3) ;
          \vertex (x2) at ( 1/3, -1/3) ;
          \vertex (y) at (1.5, 0) ;
          \vertex (y1) at (2.5, 0) ;
          \vertex (a) at (3, 1) ;
          \vertex (b) at (3,-1); 
          \vertex[left= 0.6cm of x] (p);
          \diagram*{ 
              (p) --[scalar] (x), 
              (x) -- (x1);
              (x) -- (x2);
              (x1) --[fermion,out=45, in=120] (y),
              (y) --[fermion,out=-120, in=-45] (x2),
              (y) --[photon] (y1),  
              (y1) --[fermion] (a),
              (b) --[fermion] (y1),
          };
      \end{feynman} 
      \node[xshift=-10mm]    {\large +};
    \end{scope}
    \begin{scope}[yshift=-25mm,xshift=70mm,
      line width=1pt,scale=0.6]
      \path[name path=C]  (1/3,1/3) -- (0,0) -- (1/3,-1/3);  
      \path[name path=X]  (1/3,-1/3)  (1/3,1/3);  
      \tikzfillbetween[of=C and X]{gray!90};
      \begin{feynman}
          \vertex (x) at ( 0, 0) ;
          \vertex (x1) at ( 1/3, 1/3) ;
          \vertex (x2) at ( 1/3, -1/3) ;
          \vertex (y) at (1.5, 0) ;
          \vertex (y1) at (2.5, 0) ;
          \vertex (a) at (3, 1) ;
          \vertex (b) at (3,-1); 
          \vertex[left= 0.6cm of x] (p);
          \diagram*{ 
              (p) --[scalar,style=red] (x), 
              (x) --[style=red] (x1);
              (x) --[style=red] (x2);
              (x1) --[fermion,out=45, in=120,style=red] (y),
              (y) --[fermion,out=-120, in=-45,style=red] (x2),
              (y) --[photon] (y1),  
              (y1) --[fermion] (a),
              (b) --[fermion] (y1),
          };
      \end{feynman} 
      \node[xshift=-10mm]    {\large +};
    \end{scope}
    \begin{scope}[yshift=0mm,xshift=130mm,
      line width=1pt,scale=0.6]
      \path[name path=C]  (1/3,1/3) -- (0,0) -- (1/3,-1/3);  
      \path[name path=X]  (1/3,-1/3)  (1/3,1/3);  
      \tikzfillbetween[of=C and X]{gray!90};
      \begin{feynman}
          \vertex (x) at ( 0, 0) ;
          \vertex (x1) at ( 1/3, 1/3) ;
          \vertex (x2) at ( 1/3, -1/3) ;
          \vertex (y) at (1.5, 0) ;
          \vertex (y1) at (2.5, 0) ;
          \vertex (a) at (3, 1) ;
          \vertex (b) at (3,-1); 
          \vertex[left= 0.6cm of x] (p);
          \diagram*{ 
              (p) --[scalar] (x), 
              (x) -- (x1);
              (x) -- (x2);
              (x1) --[charged scalar,out=45, in=120] (y),
              (y) --[charged scalar,out=-120, in=-45] (x2),
              (y) --[photon] (y1),  
              (y1) --[fermion] (a),
              (b) --[fermion] (y1),
          };
      \end{feynman} 
      \node[xshift=-10mm]    {\large +};
    \end{scope}
    
    \begin{scope}[yshift=-25mm,xshift=100mm,
      line width=1pt,scale=0.6]
      \path[name path=C]  (1/3,1/3) -- (0,0) -- (1/3,-1/3);  
      \path[name path=X]  (1/3,-1/3)  (1/3,1/3);  
      \tikzfillbetween[of=C and X]{gray!90};
      \begin{feynman}
          \vertex (x) at ( 0, 0) ;
          \vertex (x1) at ( 1/3, 1/3) ;
          \vertex (x2) at ( 1/3, -1/3) ;
          \vertex (y) at (1.5, 0) ;
          \vertex (y1) at (2.5, 0) ;
          \vertex (a) at (3, 1) ;
          \vertex (b) at (3,-1); 
          \vertex[left= 0.6cm of x] (p);
          \diagram*{ 
              (p) --[scalar,style=red] (x), 
              (x) --[style=red] (x1);
              (x) --[style=red] (x2);
              (x1) --[charged scalar,out=45, in=120,style=red] (y),
              (y) --[charged scalar,out=-120, in=-45,style=red] (x2),
              (y) --[photon] (y1),  
              (y1) --[fermion] (a),
              (b) --[fermion] (y1),
          };
      \end{feynman} 
      \node[xshift=-10mm]    {\large +};
    \end{scope}

    \begin{scope}[yshift=-50mm,xshift=0mm,
      line width=1pt]
      \path[name path=C]  (1/3,1/3) -- (0,0) -- (1/3,-1/3);  
      \path[name path=X]  (1/3,-1/3)  (1/3,1/3);  
      \tikzfillbetween[of=C and X]{gray!90};
      \begin{feynman}
        \vertex (x) at ( 0, 0) ;
        \vertex (x1) at ( 1/3, 1/3) ;
        \vertex (x2) at ( 1/3, -1/3) ;
        \vertex (y) at (1.5, 0) ;
        \vertex (y1) at (2,0, 0) ; 
        \vertex[left= 0.6cm of x] (p);
        \diagram*{ 
            (p) --[scalar] (x), 
            (x) -- (x1);
            (x) -- [style=red](x2);
            (x1) --[fermion,out=45, in=120] (y),
            (y) --[fermion,out=-120, in=-45,style=red] (x2), 
            (y) --[scalar] (y1), 
        };
      \end{feynman}
      \node[below=10mm,xshift=10mm]    {(b)};
    \end{scope}
    \begin{scope}[yshift=-50mm,xshift=30mm,
      line width=1pt,scale=0.6]
      \path[name path=C]  (1/3,1/3) -- (0,0) -- (1/3,-1/3);  
      \path[name path=X]  (1/3,-1/3)  (1/3,1/3);  
      \tikzfillbetween[of=C and X]{gray!90};
      \begin{feynman}
          \vertex (x) at (0, 0) ;
          \vertex[] (a) at (1, 1) ;
          \vertex[] (b) at (1,-1); 
          \vertex[left= 0.6cm of x] (p);
          \diagram*{ 
              (p) --[scalar] (x), 
              (x) --[fermion] (a),
              (b) --[fermion,style=red] (x), 
          };
      \end{feynman}
      \node[xshift=10mm]    {\large =};
    \end{scope}
    \begin{scope}[yshift=-50mm,xshift=50mm,
      line width=1pt,scale=0.6] 
      \begin{feynman}
          \vertex (x) at ( 0, 0) ;
          \vertex[] (a) at (1, 1) ;
          \vertex[] (b) at (1,-1); 
          \vertex[left= 0.6cm of x] (p);
          \diagram*{ 
              (p) --[scalar] (x), 
              (x) --[fermion] (a),
              (b) --[fermion,style=red] (x), 
          };
      \end{feynman}
      \node[xshift=10mm]    {\large +};
    \end{scope}
    \begin{scope}[yshift=-50mm,xshift=70mm,
      line width=1pt,scale=0.6]
      \path[name path=C]  (1/3,1/3) -- (0,0) -- (1/3,-1/3);  
      \path[name path=X]  (1/3,-1/3)  (1/3,1/3);  
      \tikzfillbetween[of=C and X]{gray!90};
      \begin{feynman}
          \vertex (x) at ( 0, 0) ;
          \vertex[] (a) at (1, 1) ;
          \vertex[] (b) at (1,-1);
          \vertex[right=1cm of a] (a1);
          \vertex[right=1cm of b] (b1);
          \vertex[left= 0.6cm of x] (p);
          \diagram*{ 
              (p) --[scalar] (x), 
              (x) --[fermion] (a),
              (a) --[fermion] (a1),
              (b) --[fermion,style=red] (x), 
              (b1) --[fermion,style=red] (b),
              (a) --[photon] (b),
          };
      \end{feynman}
      \node[xshift=-10mm]    {\large +};
    \end{scope}
    \begin{scope}[yshift=-50mm,xshift=100mm,
      line width=1pt,scale=0.6]
      \path[name path=C]  (1/3,1/3) -- (0,0) -- (1/3,-1/3);  
      \path[name path=X]  (1/3,-1/3)  (1/3,1/3);  
      \tikzfillbetween[of=C and X]{gray!90};
      \begin{feynman}
          \vertex (x) at ( 0, 0) ;
          \vertex (x1) at ( 1/3, 1/3) ;
          \vertex (x2) at ( 1/3, -1/3) ;
          \vertex (y) at (1.5, 0) ;
          \vertex (y1) at (2.5, 0) ;
          \vertex (a) at (3, 1) ;
          \vertex (b) at (3,-1); 
          \vertex[left= 0.6cm of x] (p);
          \diagram*{ 
              (p) --[scalar] (x), 
              (x) -- (x1);
              (x) --[style=red] (x2);
              (x1) --[charged scalar,out=45, in=120] (y),
              (y) --[charged scalar,out=-120, in=-45,style=red] (x2),
              (y) --[photon] (y1),  
              (y1) --[fermion] (a),
              (b) --[fermion,style=red] (y1),
          };
      \end{feynman} 
      \node[xshift=-10mm]    {\large +};
    \end{scope} 
  \end{tikzpicture}
  \caption{Diagrammatic representation of the energy correction in second order of SOC strength. 
  Where the solid line represents Green's function of $\tau$ valley, 
  dashed line represents the $-\tau$ valley, black represents spin $s$ and red for spin $-s$. The top plane shows the diagrams for LAF-z state and bottom plane for LAF-x state.}
  \label{fig:feyn_semi}
\end{figure}

For the LAF-z state, the dressed vertex is given by 
\begin{equation}
  \tilde{\lambda}^z_{\tau,s}=\lambda_{\tau,s}+U\tilde{\lambda}^z_{\tau,s}\Pi_{\tau,s}-U\sum_{\tau',s'=\pm 1} \tilde{\lambda}^z_{\tau',s'}\Pi_{\tau',s'}-
  \sum_{s'=\pm 1}\tilde{\lambda}^z_{-\tau,s'}(V-J_H s^z_{s',s'}s^z_{s,s})\Pi_{-\tau,s'}
  \label{eq:LAFz_vertex}
\end{equation}
we notice the first summation is over flavors not equal to $(\tau,s)$. And the particle-hole bubble is defined as 
\begin{equation}
  \Pi_{\tau,s}=-\int \frac{d^2k}{(2\pi)^2} \sum_{n,m}\frac{u'_{n}(k)u_{m}(k)u'_{m}(k)u_{n}(k)}{E_{n,k}-E_{m,k}}(f(E_{n,k})-f(E_{m,k}))
  \label{eq:LAFz_bubble}
\end{equation}
where $u_{n}(k)$ is the wave function at the bottom layer. In the LAF-z state, where the spin and valley are conserved, we omit the spin and 
valley indices in the above formula. However, when considering the LAF-x state, instead of focusing on a state with in-plane magnetic order, 
we can rotate the Ising SOC to align with the in-plane direction, such as the x-direction. This rotation
changes the SOC term to $\lambda \tau_z s_x (\rho_0-\rho_z)/2$, allowing us to continue our analysis within the LAF-z state. Accordingly, the dressed 
vertex for the LAF-x state is 
\begin{equation}
  \tilde{\lambda}^{x}_{\tau,(s,-s)}=\lambda_{\tau,(s,-s)}+\tilde{\lambda}^{x}_{\tau,(s,-s)}U\Pi_{\tau,(s,-s)}
  +J_H \tilde{\lambda}^{x}_{-\tau,(s,-s)}\Pi_{-\tau,(s,-s)}
  \label{eq:LAFx_vertex}
\end{equation}
with 
\begin{equation}
  \Pi_{\tau,(s,-s)}=-\int \frac{d^2k}{(2\pi)^2} \sum_{n,m}\frac{u'_{n,s}(k)u_{m,-s}(k)u'_{m,-s}(k)u_{n,s}(k)}{E_{n,s,k}-E_{m,-s,k}}(f(E_{n,s,k})-f(E_{m,-s,k}))
  \label{eq:LAFx_bubble}
\end{equation}

Before presenting the numerical results, it is instructive to first analyze the structure of the polarization bubble given by Eq.~(\ref{eq:LAFz_bubble})
and Eq.~(\ref{eq:LAFx_bubble}). In the vicinity of the semi-metal region, as discussed in the previous section, the bands at the Fermi level are spin polarized. 
For instance, the $(K\uparrow,K'\uparrow)$ band becomes semi-metallic, while the $(K\downarrow,K'\downarrow)$ band exhibits a large band gap. Therefore, in the 
case of the LAF-x state, which involves bands from different spin of the same valley, the bubble described by Eq.~(\ref{eq:LAFx_bubble}) 
would be suppressed by the band gap of opposite spin. On the other hand, for the LAF-z state, contributions from the Fermi surface are significant in the semi-metal 
region, these contributions are proportional to the density of states. Notably, the pentalayer system has a much larger DOS compared to the tetralayer 
system. As a result, when the polarization bubble becomes sufficiently large, it suggests the breakdown of the perturbation approach and indicates the spontaneous
generation of an additional mass term. More precisely, Eq.~(\ref{eq:LAFz_vertex}) and Eq.~(\ref{eq:LAFx_vertex}) can be written into matrix form as 
$\tilde{\lambda}=\lambda+\hat{\Gamma}\tilde{\lambda}$. Whenever the largest eigenvalue $\Gamma_i$ of $\hat{\Gamma}$, 
with the corresponding eigenvector share a non-vanishing overlap with Ising-SOC vector, exceeds $1$, it implies the current band 
is not stable against spontaneous mass term generation. Additionally, the second order correction 
to the ground state energy is given by $\delta E=-\lambda^\top \hat{\Pi}(I-\hat{\Gamma})^{-1}\lambda/2$.
Here, $\lambda_{\text{LAF-z}}=(1,-1,-1,1)^\top$ and $\lambda_{\text{LAF-x}}=(1,-1)^\top$.

\begin{figure}[htb]
  \begin{center}
    \includegraphics[width=0.9\textwidth]{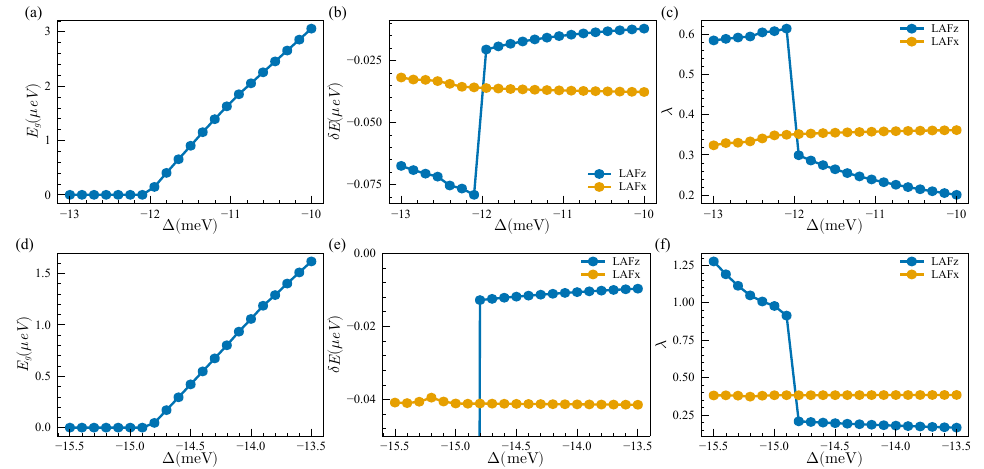}
  \end{center}
  \caption{Hartree-Fock results for the LAF state in the tetralayer and pentalayer systems as a function of the external displacement field, in the absence of SOC.
  The top panel shows the results for the tetralayer, and the bottom panel represents the pentalayer system.
  (a,d) The indirect band gap of the Hartree-Fock band structure.
  (b,e) The correction of ground state energy per unit cell due to SOC, calculated using perturbation theory. The SOC strength is 
  $\lambda=1\text{meV}$. (c,f) The largest eigenvalue of corresponding coupling matrix  . 
  Other parameters are $U= 40~\text{eV}\cdot\mathcal{A}_u$, $V=-U/5$, and $J_H=10~\text{eV}\cdot\mathcal{A}_u$.}
  \label{fig:InsulatorToSM_SOC}
\end{figure}

Fig.~\ref{fig:InsulatorToSM_SOC} shows the calculated second order ground state energy corrections for LAF state. In the gapped region, the LAF-x 
exhibits lower energy than the LAF-z state. Interestingly, the energy difference obtained from the perturbation theory closely aligns with the values 
derived from self-consistent Hartree-Fock theory. Upon entering the semi-metal region, as discussed earlier, the Fermi surface 
contributions significantly enhance the LAF-z polarization bubble, making the energy of the LAF-z state lower than that of the LAF-x state. 
For the pentalayer system, instability arises when $\Gamma_i>1$ 
consistent with the results showing that, even without external SOC, an anomalous semi-metal state has smaller energy than the normal semi-metal state in the semi-metal region. 
In contrast, for tetralayer system, the LAF-z state remains stable against interaction, which also aligns with the findings from 
the previous section. Finally, it is noted that SOC decreases the band gap of the LAF-z state while increasing the band gap of the LAF-x state. Consistently, the phase transition 
from LAF-x to LAF-z state occurs when LAF-x retains a band gap while the LAF-z has already become gapless. 
This observation explains the numerical results in Fig.~\ref{fig:InsulatorToSM_wSOC}.

\section{Derivation of inter-valley Hund's Coupling}
As discussed in the main text and the previous section, the inter-valley interaction plays a crucial role in selecting the LAF state as the ground state in absence of an external 
displacement field. Additionally, throughout our calculation, we have used a negative $V$ term. The $V$ term and Hund's coupling term can be 
derived from short-range interactions~\cite{chatterjee2022}. To justify our choose of parameters, we now present the derivation of the inter-valley Hund's coupling from atomic Hubbard term.

The atomic Hubbard term is 
\begin{equation}
  H_{\text{Hubbard}}=U_0\sum_i n_{i,\uparrow}n_{i,\downarrow}
\end{equation}
where the summation is over all the sublattices. The creation and annihilation operator can expanded as 
\begin{equation}
  c_{i,s}=\frac{1}{\sqrt{N}}\sum_{\mathbf{k}}e^{i\mathbf{k}\cdot\mathbf{r_i}} c_{\mathbf{k},s}
  =\frac{1}{\sqrt{N}}\sum_{\tau,|\mathbf{k}-\tau\mathbf{K}|<\Lambda}e^{i(\mathbf{k}-\tau\mathbf{K})\cdot\mathbf{r_i}}
  e^{i\tau\mathbf{K}\cdot\mathbf{r_i}}c_{\mathbf{k},s}+\cdots\approx\sum_{\tau}e^{i\tau \mathbf{K}}c_{i,\tau,s}
\end{equation}
where $N$ is number of unit cell, and $s=\uparrow/\downarrow$ is spin index and $\tau=\pm 1$ is valley index.
And $\mathbf{K}$ is the vector of $K$ point. The $\cdots$ represents operator with momentums outside the UV cut-off which can be omitted.
Then, the density operator is 
\begin{equation}
  \begin{aligned} 
  n_i&=\sum_s n_{i,s}=\sum_s c^\dagger_{i,s}c_{i,s}\approx\sum_{s,\tau,\tau'} c^\dagger_{i,\tau,s}c_{i,\tau',s}e^{i(-\tau+\tau')\mathbf{K}\cdot\mathbf{r}_i}\\ 
  &=\sum_{\tau,s} \tilde{n}_{i,\tau,s}+\sum_{\tau,s}c^\dagger_{i,\tau,s}c_{i,-\tau,s}e^{-2\tau i\mathbf{K}\cdot\mathbf{r}_i}\\
  \end{aligned}
\end{equation}
where $\tilde{n}_{i,\tau,s}=c^\dagger_{i,\tau,s}c_{i,\tau,s}$ is density operator for $\tau$-th valley.
Then, we notice the Hubbard term can be written as 
\begin{equation}
  \begin{aligned}
    H_{Hubbard}&=U_0\sum_i n_{i,\uparrow}n_{i,\downarrow}=\frac{U_0}{2}\sum_i (\sum_{s}n_{i,s})^2\\ 
    &\approx  \frac{U_0}{2}\sum_i (\sum_{\tau,s}\tilde{n}_{i,\tau,s}+c^\dagger_{i,\tau,s}c_{i,-\tau,s}e^{-2\tau i\mathbf{K}\cdot\mathbf{r}_i})^2\\ 
    &\approx \frac{U_0}{2} \sum_i (\sum_{\tau,s}\tilde{n}_{i,\tau,s})^2+\frac{U_0}{2} \sum_i (\sum_{\tau,s}c^\dagger_{i,\tau,s}c_{i,-\tau,s}e^{-2\tau i\mathbf{K}\cdot\mathbf{r}_i})^2\\ 
    &\approx \frac{U_0}{2} \sum_i (\sum_{\tau,s}\tilde{n}_{i,\tau,s})^2+\frac{U_0}{2}\sum_i\sum_{\tau,s}c^\dagger_{i,\tau,s}c_{i,-\tau,s}c^\dagger_{i,-\tau,s'}c_{i,\tau,s'}
  \end{aligned}
  \label{eq:HubbardTerm_expansion}
\end{equation}
where, in above formulas, we have omitted the terms which carry non-trivial momentum factor. The first term in above formula 
will contribute to the first term in Eq.~(\ref{eq:ham_int_2part}). While, the second term can be formulated as 
\begin{equation}
  \begin{aligned}
    \frac{U_0}{2}\sum_i\sum_{\tau,s}c^\dagger_{i,\tau,s}c_{i,-\tau,s}c^\dagger_{i,-\tau,s'}c_{i,\tau,s'}&=
    -\frac{U_0}{2}\sum_i\sum_{\tau,s}c^\dagger_{i,\tau,s}c_{i,\tau,s'}c^\dagger_{i,-\tau,s'}c_{i,-\tau,s}\\ 
    &=-\frac{U_0}{2}\sum_i\sum_{\tau,\alpha,\beta,\gamma,\eta}
    c^\dagger_{i,\tau,\alpha}c_{i,\tau,\beta}c^\dagger_{i,-\tau,\gamma}c_{i,-\tau,\eta}\delta_{\alpha,\eta}\delta_{\beta,\gamma}\\ 
    \label{eq:HubbardTerm_expansion2}
  \end{aligned}
\end{equation}
then, with Fierz identity $\delta_{\alpha,\beta}\delta_{\gamma,\eta}+\vec{\sigma}_{\alpha,\beta}\cdot\vec{\sigma}_{\gamma,\eta}
=2\delta_{\alpha,\eta}\delta_{\beta,\gamma}$, we have 
\begin{equation}
  \begin{aligned}
    \text{Eq.}~(\ref{eq:HubbardTerm_expansion2})&=-\frac{U_0}{4}\sum_i\sum_{\tau,\alpha,\beta,\gamma,\eta}
    c^\dagger_{i,\tau,\alpha}c_{i,\tau,\beta}c^\dagger_{i,-\tau,\gamma}c_{i,-\tau,\eta}(\delta_{\alpha,\beta}\delta_{\gamma,\eta}+\vec{\sigma}_{\alpha,\beta}\cdot\vec{\sigma}_{\gamma,\eta})\\
    &=-\frac{U_0}{4}\sum_{i,\tau}\tilde{n}_{i,\tau}\tilde{n}_{i,-\tau}+\mathbf{S}_{i,\tau}\cdot\mathbf{S}_{i,-\tau}
    =-\frac{U_0}{2}\sum_{i}\tilde{n}_{i,K}\tilde{n}_{i,K'}+\mathbf{S}_{i,K}\cdot\mathbf{S}_{i,K'}\\ 
    &=-\bar{U}_0\int d^2x \sum_{l} \tilde{n}_{l,K}(\mathbf{x})\tilde{n}_{l,K'}(\mathbf{x})+
    \mathbf{S}_{l,K}(\mathbf{x})\cdot\mathbf{S}_{l,K'}(\mathbf{x})\\ 
  \end{aligned}
\end{equation}
where the summation in last line means sum over layers and $\bar{U}=(\sqrt{3}a^2)U_0$. We have dropped the constant chemical potential term 
from anticommutation relation of fermionic operators.
In summary, starting from the atomic Hubbard term, 
we can indeed derive a ferromagnetic inter-valley Hund's coupling and a negative $V$ term. It is important to note that in the above derivation, 
all three interaction terms share the same amplitude. However, in our calculations, the introduction of a UV cut-off implies 
our model should be viewed as a low energy effective theory obtained through renormalization group by integrating out high energy degrees 
of freedom. Therefore, we expect through the coefficients to differ from one another.

Finally, we note that another form of Hund's coupling, $\mathbf{S}_{+,-}\cdot \mathbf{S}_{-,+}$, where 
$\mathbf{S}_{+,-}=c^\dagger_{K}\vec{\sigma}c_{K'}$, is widely used in the literature~\cite{chatterjee2022,you2022}. This term can be related to ours through the Fierz identity. For example, the 
Eq.~(\ref{eq:HubbardTerm_expansion2}) can be written as 
\begin{equation}
  \begin{aligned}
    \frac{U_0}{2}\sum_i\sum_{\tau,s}c^\dagger_{i,\tau,s}c_{i,-\tau,s}c^\dagger_{i,-\tau,s'}c_{i,\tau,s'}&=
    \frac{U_0}{2}\sum_i\sum_{\tau}c^\dagger_{i,\tau,\alpha}c_{i,-\tau,\beta}c^\dagger_{i,-\tau,\gamma}c_{i,\tau,\eta}\delta_{\alpha,\beta}\delta_{\gamma,\eta}\\ 
    &=\frac{U_0}{2}\sum_i\sum_{\tau}c^\dagger_{i,\tau,\alpha}c_{i,-\tau,\beta}c^\dagger_{i,-\tau,\gamma}c_{i,\tau,\eta}
    (2\delta_{\alpha,\eta}\delta_{\beta,\gamma}-\vec{\sigma}_{\alpha,\beta}\cdot\vec{\sigma}_{\gamma,\eta})\\ 
    &=-U_0\sum_i (2n_{K}n_{K'}+\mathbf{S}_{+,-}\cdot \mathbf{S}_{-,+})
  \end{aligned}
  \label{eq:valleyHundCoupling}
\end{equation}
Even though those two different formulas are equivalent to each other, we still use the spin Hund's coupling in our calculation.
\section{Charge Density Distribution}
In the main text, the interlayer potential $\Delta$ is related to the external displacement field through a simple linear relation.
Interestingly, we found that the 
critical field for the LAF to SM transition aligns well with experimental observations. However, the field for the SM to LAF transition field deviates 
significantly from experimental results.
To provide a theoretical understanding of this discrepancy, we present the charge density distribution in this section. 

\begin{figure}[htb]
  \begin{center}
    \includegraphics[width=0.8\textwidth]{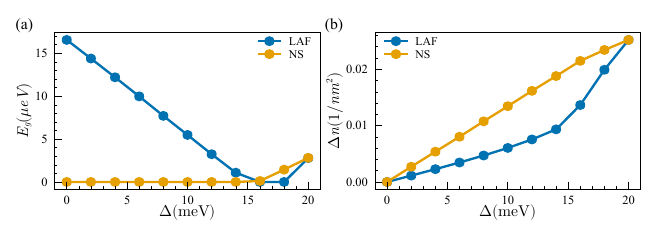}
  \end{center}
  \caption{Hartree-Fock results for density distribution of pentalayer system as a function of the external displacement field in absence of SOC. 
  The "NS" refers to the non-symmetry breaking normal state.
  (a) The indirect band gap of the LAF and normal states.
  (b) The density difference between bottom layer and top layer. The 
  parameters are $U= 40~\text{eV}\cdot\mathcal{A}_u$, $V=-U/5$, and $J_H=10~\text{eV}\cdot\mathcal{A}_u$.}
  \label{fig:density}
\end{figure}

Fig.~\ref{fig:density} shows the charge density distribution as a function of the interplay potential $\Delta$. As discussed in the main text,
the LAF state exhibits vanishing layer polarization at $\Delta=0$, resulting in a weak response to the potential $\Delta$. 
Our calculations indicate that when $\Delta$ is small and the system remains in the LAF state, 
the charge density is almost uniformly distributed across the system. Only upon entering the SM phase does
the charge density significantly deviate, indicating a strong screening effect.

After obtaining the charge density distribution, we can proceed to calculate the screened interlayer potential following Refs.~\cite{Koshino2009,koshino2010}.
However, since our model omits interlayer interactions, part of the screening effect is already accounted for.
Therefore, an additional Possion's equation calculation could introduce a double-counting problem. Consequently, we expect that, at least in the LAF state, 
the $\Delta-D$ relation will approximately follow the simple linear relation presented in the main text.

\section{Chern Number}
In this section, we discuss how to calculate the Chern number in rhombohedral graphene. We consider only the states with 
spin $U(1)_s$ and valley $U(1)_v$ symmetries. In this case, the total Chern number is given by the sum of the Chern numbers for each of the four degrees of freedom.
The Chern number for a single flavour can be obtained using the TKNN formula. Alternatively, the Chern number can be determined by counting the 
number of band inversions as the system transitions from a trivial insulator to a Chern insulator~\cite{Bernevig2013}.

For the first method, we only need to focus on one degree of freedom. Using the tetralayer system as 
an example, the Hamiltonian is given by Eq.~(\ref{eq:ham_ABC}) with an additional mass term. However, the complexity of the Hamiltonian prevents an analytical 
calculation of the Chern number. To simplify the problem, we notice the low energy physics in rhombohedral graphene consist of top A sublattice and 
bottom B sublattice. Therefore, in a gapped system, the Hamiltonian can be projected onto 
this low energy subspace but without affecting the band topology. According to Ref.~\cite{ghazaryan2023,koshino2009a,zhang2010a}, the effective Hamiltonian is given by 
\begin{equation}
  H_{\text{eff}}= v_{\text{eff}}k^N(\cos{N\varphi_{k}}\rho_x+\sin{N\varphi_{k}}\rho_y)+m\rho_z+\dots=\sum_{\alpha=1}^3 d_{\alpha}(\mathbf{k})\rho_\alpha+\cdots
  \label{eq:effectiveH}
\end{equation}
where $v_{\text{eff}}=v^N_0/(-\gamma)^{N-1}_1$, the $\dots$ representing the high-order terms which are irrelevant for determining the topology of a fully gapped band, 
and $\cos{\varphi_k}=\tau k_x/k, \sin{\varphi_k}=k_y/k$. For this effective Hamiltonian, one can obtain the Chern number through TKNN formula\cite{Bernevig2013}
\begin{equation}
  C=\frac{1}{4\pi}\int d^2k \frac{(\partial_{k_x}\mathbf{d}\times \partial_{k_y}\mathbf{d})\cdot \mathbf{d}}{d^3}
  =\frac{1}{4\pi}\int^{\infty}_{0} 2\pi kdk \frac{N^2 \tau m k^{2(N-1)}}{m^2+k^{2N}}=\frac{N\text{sign}(m)\tau}{2}
  \label{eq:chern_integral}
\end{equation}
therefore, the total Chern number is $(\text{sign}(m_{K,\uparrow})+\text{sign}(m_{K,\downarrow})-\text{sign}(m_{K',\uparrow})-\text{sign}(m_{K',\downarrow}))N/2$.

However, one might concern that the effective model is only well-defined around the $K$ and $K'$ points, while in Eq.~(\ref{eq:chern_integral}) we need to integrate to 
infinity. To address this possible confusion, we can consider the change in the Chern number when system undergoes a gap-closing and reopening process. If we treat 
the mass term $m_z$ as a third-dimensional momentum, i.e. $k_z$, then, the $\mathbf{k}=0$ point can be view as a three dimensional Weyl point. Consequently, the 
change in Chern number is related to the Chern number of the Weyl point as 
\begin{equation}
  C(m_{\tau,z}>0)-C(m_{\tau,z}<0)=\tau C_{\text{Weyl}}
  \label{eq:chern_change}
\end{equation}
To calculate the Chern number of the Weyl point, we can still keep the leading term in the effective Hamiltonian Eq.~(\ref{eq:effectiveH}), this can be 
achieved by choosing a large $|k|$ while within the validity of the effective model. Parameterize the $\mathbf{d}$ as 
$\mathbf{d}=\sqrt{k^{2N}+m_z^2}(k^N\cos N\varphi,k^N\sin N\varphi,m_z)/\sqrt{k^{2N}+m_z^2}=
|d|(\sin{\theta} \cos{N\varphi},\sin{\theta}\sin{N\varphi},\cos{\theta})$ (set $v_{eff}=1$), the corresponding eigenstates with energy $\pm |d|$ are 
\begin{equation}
  |-,\theta,\phi\rangle=\left(\begin{array}{c}
      \sin(\theta/2) e^{-iN\varphi} \\ 
      -\cos(\theta/2)  \\ 
  \end{array}
  \right)\quad 
  |+,\theta,\phi\rangle=\left(\begin{array}{c}
      \cos(\theta/2) e^{-iN\varphi} \\ 
      \sin(\theta/2)  \\ 
  \end{array}
  \right)
\end{equation}
then the corresponding Berry potential for $|-,\theta,\phi\rangle$ is 
\begin{equation}
  A_{\theta}=i\langle -,\theta,\varphi |\partial_{\theta} | -,\theta,\varphi \rangle=0
  \quad A_{\varphi}=i\langle -,\theta,\varphi |\partial_{\varphi} | -,\theta,\varphi \rangle=N \sin^2(\theta/2)
\end{equation}
And the Berry curvature is 
\begin{equation}
  F_{\theta,\varphi}=\partial_{\theta}A_{\varphi}-\partial_{\varphi}A_{\theta}=N\sin(\theta)/2
\end{equation}
as 
\begin{equation}
  \theta=\arcsin \frac{(k_x^2+k_y^2)^{N/2}}{\sqrt{(k_x^2+k_y^2)^N+k_z^2}}
  \quad 
  \varphi=\arccos \frac{k_x}{\sqrt{(k_x^2+k_y^2)}}
\end{equation}
in the momentum coordinate, the Berry curvature is 
\begin{equation}
  F_{k_i,k_j}=F_{\theta,\varphi}\frac{\partial(\theta,\varphi)}{\partial (k_i,k_j)}
\end{equation}
after explicit calculation, we have 
\begin{equation}
  F_{k_x,k_y}=\frac{N^2k^{2(N-1)} k_z}{2 d^3}\quad 
  F_{k_x,k_z}=-\frac{Nk^{2(N-1)} k_y}{2 d^3}\quad
  F_{k_y,k_z}=\frac{Nk^{2(N-1)} k_x}{2 d^3}
\end{equation}
the Berry field strength $V_i=\epsilon_{ijk}F_{k_i,k_j}/2$ is \cite{Bernevig2013}:
\begin{equation}
  \mathbf{V}=(k_x,k_y,N k_z)\frac{N k^{2(N-1)}}{2d^3}
\end{equation}
where $k^2=k_x^2+k_y^2$, $d^2=k^{2N}+k_z^2$.
And the Chern number of the Weyl point as 
\begin{equation}
  C_{\text{Weyl}}=\frac{1}{2\pi}\oint d\mathbf{S}\cdot \mathbf{V}=N
\end{equation}
If we further considering the fact, due to time-reversal symmetry, the layer polarized state ($m_{K,\uparrow\downarrow}=m_{K',\uparrow\downarrow}$) has zero Chern number. Then, 
the Chern number of general state could be obtained through Eq.~(\ref{eq:chern_change}). 
The finial result is the same as obtained from the TKNN formula.

%